\documentclass[conference,letterpaper]{IEEEtran}

\usepackage[utf8]{inputenc} 
\usepackage[T1]{fontenc}
\usepackage{ifthen}
\usepackage{cite}
\usepackage[cmex10]{amsmath} % Use the [cmex10] option to ensure complicance
\usepackage{amsmath}  % Define \boldsymbol (in amsbsy too) and align
\usepackage{amssymb}  % Define \mathbb (in amsfonts too)
\usepackage{mathrsfs} % Define \mathscr, a script font
\usepackage{bbm}

\usepackage{multirow}
\usepackage{subfigure}

\usepackage[noend]{algorithmic}
\usepackage{algorithm}

\usepackage{theorem}  % Helps in rendering theorems, etc.
\usepackage{comment}  % Comments out with \begin{comment} ... \end{comment}

\usepackage{upref}
\usepackage{amsfonts}

\usepackage{verbatim}

\usepackage[numbers,sort&compress]{natbib}

\usepackage[dvipsnames,usenames]{color}

\usepackage{graphicx}
\usepackage{subcaption}
\usepackage{tikz}

\usepackage{latexsym}
\usepackage{makecell}
\usepackage{enumitem}

\newtheorem{theorem}{Theorem}%[section]
\newtheorem{cor}{Corollary}
\newtheorem{lemma}{Lemma}

\newtheorem{definition}{Definition}

\usepackage{pgfplots}
\usepgfplotslibrary{dateplot}
\pgfplotsset{compat=1.8}

\usepackage{xurl}
\usepackage[colorlinks=true, linkcolor=black, urlcolor=black, citecolor=black]{hyperref}

\begin{document}

%
% \newcommand\relatedversion{}
% \renewcommand\relatedversion{\thanks{The full version of the paper can be accessed at \protect\url{https://arxiv.org/abs/1902.09310}}} % Replace URL with link to full paper or comment out this line

%\setcounter{chapter}{2} % If you are doing your chapter as chapter one,
%\setcounter{section}{3} % comment these two lines out.

%\title{\Large Guaranteed Recovery of Sufficiently Unambiguous Clusters}%\relatedversion}
% \author{Kayvon Mazooji\thanks{University of Illinois Urbana-Champaign.}
% \and Ilan Shomorony\thanks{University of Illinois Urbana-Champaign.}}

\title{Guaranteed Recovery of Unambiguous Clusters} 

% %%% Single author, or several authors with same affiliation:
\author{%
 \IEEEauthorblockN{Kayvon Mazooji and Ilan Shomorony}
\IEEEauthorblockA{Electrical and Computer Engineering\\
                   University of Illinois Urbana-Champaign\\
                   Urbana IL, United States\\
                   Email: mazooji2@illinois.edu, ilans@illinois.edu}}

%%% Several authors with up to three affiliations:
% \author{%
%   \IEEEauthorblockN{Author 1}
%   \IEEEauthorblockA{Department of Electrical Engineering \\
%                     University 1\\
%                     City 1\\
%                     Email: author1@university1.edu}
%   \and
%   \IEEEauthorblockN{Author 2 and Author 3}
%   \IEEEauthorblockA{Research Center XY\\ 
%                     City 2\\
%                     Email: \{author2, author3\}@research-center.com}
% }

%\date{}

\maketitle

\begin{abstract} %\small\baselineskip=9pt 
Clustering is often a challenging problem because of the inherent ambiguity in what the ``correct'' clustering should be.
Even when the number of clusters $K$ is known, this ambiguity often still exists, particularly when there is variation in density among different clusters, and clusters have multiple relatively separated regions of high density.  
In this paper we propose an information-theoretic characterization of when a $K$-clustering is ambiguous, and design an algorithm that recovers the clustering whenever it is unambiguous.
%In this paper, we introduce a new algorithm that provably recovers a $K$-clustering $C$ if it satisfies a relaxed cluster separability condition.  
This characterization formalizes the situation when two high density regions within a cluster are separable enough that they look more like two distinct clusters than two truly distinct clusters in the $K$-clustering.
The algorithm first identifies $K$ partial clusters (or ``seeds'') using a density-based approach, and then adds unclustered points to the initial $K$ partial clusters in a greedy manner to form a complete clustering.
%This framework allows for the guaranteed recovery of sufficiently unambiguous non-convex clusters with arbitrarily many relatively separated regions of high density, and arbitrary variation in density among different clusters.
We implement and test a version of the algorithm that is modified to effectively handle overlapping clusters, and observe that it requires little parameter selection and displays improved performance on many datasets compared to widely used algorithms for non-convex cluster recovery.
\end{abstract}

\section{Introduction}

%More often than not, 
There is often significant ambiguity in what the ``correct'' clustering is for a dataset.  Even in the case when the number of clusters $K$ is known, this ambiguity often still exists because a cluster can have multiple relatively separated regions of high density, and clusters can have very different densities. 
These difficulties cause many algorithms to incorrectly separate a true cluster into multiple clusters, while merging true clusters or failing to detect sufficiently prominent true clusters of low density.
These issues are compounded by the well-known fact that many clustering algorithms are ineffective at identifying non-convex clusters, which are present in many applications including image segmentation \cite{hou2016dsets}, geospatial data \cite{birant2007st, stonebraker1993sequoia}, and time series data \cite{ding2015yading}.
%To address these problems, we propose an information-theoretic condition for determining when a clustering is ambiguous, and design an algorithm that recovers the clustering whenever it is sufficiently unambiguous.

In this work, we try to overcome these difficulties from an information-theoretic perspective %in $K$-clustering 
%from a theoretical and practical standpoint 
by proposing a condition that should hold for a $K$-clustering $C$ to be considered unambiguous, along with an algorithm that recovers $C$ whenever this condition holds.
%through the lens of density-based clustering.  
%In specific, we design an efficient algorithm that provably recovers a $K$-clustering $C$ under a novel and intuitive separability condition.  
This condition characterizes the situation where two high density regions within the same cluster in $C$ look more like two distinct clusters than two truly distinct clusters in $C$.
Our framework yields the provable recovery of unambiguous $K$-clusterings that can have
\begin{itemize} %[noitemsep, topsep=0.2pt]
\item clusters with arbitrarily many relatively separated regions of high density
\item arbitrary variation in density among different clusters
\item arbitrary variation in density within clusters
\item arbitrarily shaped clusters.
\end{itemize}
Our approach is information-theoretic in the sense that our algorithm guarantees recovery of a $K$-clustering whenever the dataset warrants an unambiguous $K$-clustering in our framework.  This is in contrast to algorithms like $K$-means that are motivated by the optimization of a clustering quality measure.

%The algorithm for accomplishing this is somewhat complex

The clustering algorithm we propose is somewhat complex, but has a computationally efficient runtime.
It works by first finding a small subset of points called a ``seed'' from each cluster, and then expanding the seeds to form clusters.
We implement and test a version of the algorithm that is modified to handle overlapping clusters well, and observe that it requires little parameter selection, and delivers improved performance on artificial and benchmark datasets compared to widely used algorithms for non-convex cluster recovery.
This implementation is available at:
\url{https://github.com/kmazooji/Minimal-Seed-Expansion}.

%This is in contrast to many of the widely used density-based clustering algorithms that do not have formal recovery guarantees.
 
We begin by discussing related work and introducing background information and notation.
We then present our main theoretical results, followed by experiments comparing our algorithm to existing algorithms.  
% An implementation of the algorithm used in these experiments is available at: \url{https://github.com/kmazooji/Minimal-Seed-Expansion}.

%propose two necessary conditions for a clustering to be unambiguous.  
%Next, we design an algorithm and prove that it recovers any clustering where these conditions hold. 
%Finally, we perform experiments on a version of the algorithm that is modified to handle overlapping clusters well, and compare with other algorithms.
%We then introduce several novel cluster separability conditions of varying degrees of strictness, including our primary separability condition.  
%Next, we present our algorithm, and prove that it recovers the true clustering under each of these separability conditions.  
%Finally, we implement and test a version of the algorithm that is modified to handle overlapping clusters well, and compare its performance with widely used clustering algorithms for non-convex cluster discovery on artificial and real-world benchmark datasets.  

% \iscomment{\bf are you planning on adding a detailed section on related work? the paper currently has only one reference?}

% \kmcomment{\bf Yes I added this below.}

\section{Related Work}

The most widely used paradigms for finding non-convex clusters are spectral clustering, and density-based clustering. 
Spectral clustering algorithms perform dimensionality reduction on the data points, and run a simple clustering algorithm such as $K$-means to cluster the low dimensional data \cite{von2007tutorial}. Density-based clustering  algorithms find regions of high point density, and then output a set of high density clusters according to some criterion \cite{ester1996density, ankerst1999optics, mcinnes2017hdbscan, bhattacharjee2021survey}.
While spectral clustering is very well studied from a theoretical standpoint, most widely used density-based algorithms do not come with theoretical guarantees on when a clustering is recoverable.

The first widely-used density-based clustering algorithm was DBSCAN \cite{ester1996density}. %, which was introduced in 1996.
%, and is described in the preliminaries section. 
Since then, many algorithms have been designed to improve upon various aspects of DBSCAN \cite{ankerst1999optics, mcinnes2017hdbscan, ZHU2016, ram2010density, wang2022amd, hess2019spectacl, jang2019dbscan++, bhattacharjee2021survey}.
In addition to DBSCAN, the other most widely used density-based clustering algorithms are OPTICS \cite{ankerst1999optics} and HDBSCAN \cite{mcinnes2017hdbscan}, which were both designed to improve upon DBSCAN's ability to output clusters of varying density.  
On the theoretical side, there has been a recent line of work studying $\lambda$-density level set estimation using DBSCAN \cite{jiang2017density, jang2019dbscan++, esfandiari2021almost}.

Our algorithm works by first identifying $K$ partial clusters (or “seeds”) using a density-based approach, and then adding unclustered points to the initial $K$ partial clusters in a greedy manner to form a complete clustering.
To extract these seeds, we sequentially find seeds of decreasing density.
%$\epsilon$-partial clusters for increasing values of $\epsilon$ in a novel manner.
The intuitive idea of sequentially finding disjoint clusters of decreasing density has been employed in various work \cite{ram2010density, wang2022amd}. However, to the best of our knowledge, our algorithm for finding $K$ seeds and our algorithm for expanding the seeds to form complete clusters have not previously appeared in the literature.  
Furthermore, we are not aware of any existing mathematical analysis of density-based clustering algorithms that give guarantees similar to those presented in this work.

\section{Preliminaries}

We use $X$ to denote a set of data points.  
The distance between points $x$ and $y$ is denoted by $d(x, y)$.
As in many density-based clustering papers, the measure of density at a point $x\in X$ is determined by an integer $N_p$, and is defined as $1/\epsilon_{N_p}(x)$ where $\epsilon_{N_p}(x)$ is the minimum distance $\epsilon$ such that there are $N_p$ points in $X$ at distance at most $\epsilon$ from $x$ (including $x$ itself).  In other words, $\epsilon_{N_p}(x)$ is the distance from $x$ to its $(N_p - 1)$th nearest neighbor in $X$.
We call $\epsilon_{N_p}(x)$ the sparsity at $x$.

A cluster is simply a set of points in $X$.  A clustering $C$ of $X$ is a set of disjoint clusters, where every point in $X$ belongs to exactly one cluster (i.e. a partitioning of $X$ into clusters).  
A $K$-clustering is a clustering with $K$ clusters.
A partial clustering of $X$ is a set of disjoint clusters whose union does not necessarily include all points in $X$. We say that a clustering $C$ extends (or is an extension of) a partial clustering $C'$ if there exists a bijective function $f$ that maps the clusters in $C'$ to the clusters in $C$ such that for each cluster $c' \in C'$, $c'$ is a non-empty subset of the cluster $f(c') \in C.$ 

%To define a cluster in this model, 
We say a point $x_1$ is $\epsilon$-connected to a point $x_t$ if there exists some sequence of points $x_1, x_2, ..., x_t$ such that $x_{i+1}$ is distance at most $\epsilon$ from $x_i$ for $1 \leq i \leq t-1$ and $\epsilon_{N_p}(x_i) \leq \epsilon$ for $1 \leq i \leq t$.  %We say that 
A set of points $c$ is called $\epsilon$-connected if every pair of points in $c$ is $\epsilon$-connected.

For a given $\epsilon$, a set of points $c$ is called an $\epsilon$-cluster if it is a maximal $\epsilon$-connected set of points.  Any point that is $\epsilon$-connected to a point in an $\epsilon$-cluster $c$ is included in $c$. 
A set of points $c$ is called a maximal cluster if it is an $\epsilon$-cluster for some $\epsilon$.
For a given $\epsilon$, it is helpful to consider the graph that is formed where each node corresponds to a point in $X$, and an edge is drawn between two nodes if the corresponding points $x_1, x_2$ are such that $\epsilon_{N_p}(x_1) \leq \epsilon$, $\epsilon_{N_p}(x_2) \leq \epsilon$, and $d(x_1, x_2) \leq \epsilon$.  A set of points is an $\epsilon$-cluster if and only if it corresponds to a connected component in this graph.

The $\epsilon$-cluster centered at a point $x$ is defined as the set of points that are $\epsilon$-connected to $x$, and is denoted by $c^*(x,\epsilon)$.  If $\epsilon < \epsilon_{N_p}(x)$, then $c^*(x,\epsilon) = \{\}$.  The sparsity of a set of points $c$ is defined as the minimum $\epsilon$ such that $c$ is $\epsilon$-connected, and is denoted by $\epsilon^*(c)$.  The $\epsilon$-distance between points $x$ and $y$ is defined as the minimum $\epsilon$ such that $x$ is $\epsilon$-connected to $y$, and is denoted by $\epsilon(x, y)$.
The minimum $\epsilon$-distance from a point $x$ to a cluster $c$ is defined as $\epsilon(x, c) = \min_{y \in c} \epsilon(x, y)$. %$\epsilon(x, c) = \min_{y \in c \text{ s.t. } y \text{ is } \epsilon \text{-connected to } x} \epsilon$.
The minimum $\epsilon$-distance from cluster $c_1$ to cluster $c_2$ is defined as $\epsilon(c_1, c_2) = \min_{x \in c_1, y \in c_2} \epsilon(x, y)$.

For a dataset $X$ and density parameter $N_p$, the dendrogram $G = (V, E)$ is a tree structure that gives all $\epsilon$-clusters in $X$ for each $\epsilon \geq 0$.  $V$ and $E$ are the sets of nodes and edges in $G$ respectively.
%, and $E$ is the set of edges in $D$.  %The dendrogram is a tree where 
Each node $v \in V$ corresponds to an $\epsilon$-cluster for some $\epsilon$, and the clusters corresponding to the children of $v$ form the smallest possible partition of the maximal cluster corresponding to $v$ into maximal clusters.
Each value of $\epsilon$ specifies a
clustering given by all nodes in $V$ corresponding to $\epsilon$-clusters.  
For a given $x$ and $N_p$, the dendrogram is unique, and yields a hierarchy of possible clusterings, making it the foundational structure in hierarchical density-based clustering (HDBSCAN) \cite{mcinnes2017hdbscan}.

For a node $v \in V$, $c(v)$ denotes the cluster corresponding to $v$.  
Each node $v$ has a real number $\epsilon(v)$ associated with it, where $\epsilon(v)$ is the smallest $\epsilon$ such that $c(v)$ is a $\epsilon$-cluster.  For a leaf node $v$ corresponding to the cluster $\{x\}$, $\epsilon(v) = \epsilon_{N_p}(x)$. 
A node $v_0$ has children $v_1, v_2, ..., v_i$ if $c(v_1), c(v_2), ...,  c(v_i)$ form the smallest possible partition of $c(v_0)$ composed of maximal clusters, which is guaranteed to be unique.  
The root node $v_r$ of the resulting tree is such that $c(v_r) = X$.
%corresponds to all points in $X$. 

Any $S \subset V$ such that no node in $S$ is a descendant of another node in $S$ induces a (partial) clustering of $X$ given by $\{c(v) \text{ for } v \in S\}$.
Any such clustering is called a dendrogram clustering. 
%Any such clustering induced by the dendrogram is referred to as a dendrogram clustering.  
In fact, any (partial) clustering that consists of maximal clusters is a dendrogram (partial) clustering since every maximal cluster corresponds to a node in the dendrogram.
The $\epsilon$-cut of a dendrogram is the (partial) clustering that includes all clusters corresponding to nodes $v$ such that $\epsilon(v) \leq \epsilon$, and $v$ has no parent $v'$ with $\epsilon(v') \leq \epsilon$.  For any integer $K$, there exists at most one partial clustering given by an $\epsilon$-cut of $G$ that contains $K$ clusters.
The (partial) clustering output by DBSCAN for a given $N_p$ and $\epsilon$ can be formed by taking the $\epsilon$-cut of $G$, and adding any unclustered point $x$  %that is not part of any cluster 
to the cluster $c$ that minimizes $d(x, c)$ if $d(x, c) \le \epsilon$ where $d(x, c) = \min_{y \in c} d(x, y)$. %(ties can be broken arbitrarily)

\section{Results}

We first define two conditions that should be satisfied for a $K$-clustering $C$ of $X$ to be unambiguous.  The first condition is called weak separability, and the second is called local maximum separability.  We then design an algorithm that is guaranteed to recover $C$ if these two conditions are satisfied.

% Recovering a clustering $C$ of $X$ with a known number of clusters $K$ is often difficult because two high density regions within a cluster in $C$ can look more like two distinct clusters than two truly distinct clusters in $C$.  We introduce a novel cluster separability condition that does not allow this to happen, and design an algorithm that is guaranteed to recover $C$ when the condition is satisfied.  We begin this discussion by comparing and contrasting several notions of cluster separability.

\subsection{Weak Separability}
The simplest density-based notion of cluster separability %cluster separability %that comes to mind 
is what we call weak separability.
\begin{definition}
    For a given $N_p$, $C$ is  called weakly separable if for each $c \in C$,  $c$ is $\epsilon$-connected for some $\epsilon < \min_{c'\in C, \; c' \neq c }\epsilon(c, c')$.
\end{definition}
%If a clustering $C$ is not weakly separable, it may be clear what the true clusters roughly are, but we argue that there will be ambiguity in the exact assignment of points to clusters. 
%This is because if $C$ is not weakly separable, there must exist a cluster $c \in C$ such that any $\epsilon$-cluster containing $c$ must include at least one point from a distinct cluster $c' \in C$.

If a clustering $C$ is not weakly separable, there is ambiguity in what the correct clustering should be in the sense that there must exist a cluster $c \in C$ such that any $\epsilon$-cluster containing $c$ must include at least one point from a distinct cluster $c' \in C$.  Figures \ref{fig:sparsity_plot} and \ref{fig:weak_not_strong} illustrate weak separability.

Interestingly, a dataset $X$ often has more than one weakly separable $K$-clustering. 
For example, let $N_p = 2$ and 
$C =  [ \{1, 3, 5, 7.02, 9.02, 11.02\},$ $\{17, 18, 19, 20\},$  $\{22.01, 23.01, 24.01, 25.01\}].$
% \[C =  [ \{1, 3, 5, 7.02, 9.02, 11.02\}, \{17, 18, 19, 20\},  \{22.01, 23.01, 24.01, 25.01\}].\]
% \begin{align*}
% C =  [ &\{1, 3, 5, 7.02, 9.02, 11.02\}, \\ & \{17, 18, 19, 20\}, \{22.01, 23.01, 24.01, 25.01\}].
% \end{align*}  
Clearly, $C$ is weakly separable.  However, 
$C' = [  \{1, 3, 5\},$ $\{7.02, 9.02, 11.02\},$ $\{17, 18, 19, 20, 22.01, 23.01, 24.01, 25.01\}]$
% \[C' = [  \{1, 3, 5\}, \{7.02, 9.02, 11.02\}, \{17, 18, 19, 20, 22.01, 23.01, 24.01, 25.01\}]\]
% \begin{align*}
% C' = [ & \{1, 3, 5\}, \{7.02, 9.02, 11.02\}, \\ & \{17, 18, 19, 20, 22.01, 23.01, 24.01, 25.01\}]
% \end{align*} 
is also a weakly separable $3$-clustering.
Intuitively, $C$ is the correct clustering because the spacing between points is nearly uniform within each cluster in $C$, while there is a large relative gap in the middle of the third cluster in $C'$.
%For example, if $N_p = 2$, and $X = \{1, 2, 3, 5, 6, 7, 20, 21, 22, 26, 27, 28\}$, then $[\{1, 2, 3\}, \{5, 6, 7\}, \{20, 21, 22, 26, 27, 28\}]$ and $[\{1, 2, 3, 5, 6, 7\}, \{20, 21, 22\}, \{26, 27, 28\}]$ are both weakly separable $3$-clusterings.
It is indeed very common that the intuitively correct clustering is weakly separable, but is not the unique weakly separable clustering.
In fact, it follows from Lemma \ref{lemma:heirarchical_guarantee} that any set of maximal clusters that partition $X$ (a dendrogram clustering) is weakly separable.
\begin{lemma} \label{lemma:heirarchical_guarantee}
    For a given $N_p$, $C$ is weakly separable if and only if it is a dendrogram clustering.
\end{lemma}
\begin{IEEEproof}
    Suppose $C$ is a dendrogram clustering, but is not weakly separable.  Let $G=(V,E)$ be the dendrogram. 
    Then there exists some $v, v' \in V$ such that $c(v), c(v') \in C$, we have that $\epsilon(v) \geq \epsilon(c(v), c(v')),$  then $c(v)$ would include points from $c(v')$ since $c(v)$ is maximal. % by the definition of a dendrogram.  
    This contradicts the definition of a clustering.

    If $C$ is weakly separable, then each cluster in $C$ is maximal. 
    This follows because if a cluster $c$ is not maximal, there exists a point $x \notin c$ that is $\epsilon$-connected to $c$ such that $\epsilon \leq \epsilon^*(c)$, which contradicts weak separability.
    %implies that $c$ is not a cluster in a weakly separable clustering.
    Every maximal cluster has a corresponding node in $G$, so $C$ is a dendrogram clustering.  
\end{IEEEproof}
\begin{cor}
\label{cor:heirarchical_guarantee}
    For a given $N_p$, $X$ has a unique weakly-separable $K$-clustering if and only if exactly one dendrogram $K$-clustering exists.
\end{cor}
% \begin{IEEEproof}
%     This follows %immediately
%     from Lemma \ref{lemma:heirarchical_guarantee}.
% \end{IEEEproof} 

% Because there often does not exist a unique weakly separable $K$-clustering even when an intuitively correct $K$-clustering exists, it is not meaningful to use weak separability alone as a sufficient condition for clustering recoverability.  
If there are two weakly separable clusterings for $X$, it is impossible to guarantee that we recover one of them if our only criterion is to find a weakly separable clustering. 
Because there usually does not exist a unique weakly separable $K$-clustering even when an intuitively correct $K$-clustering exists, weak separability alone as a sufficient condition for clustering recoverability is not adequate.  %general enough.

\subsection{Local Maximum Separability}

% The main conceptual contribution of this paper is a novel cluster separability condition we call local maximum separability (or LM-separability), along with a new algorithm that provably recovers any weakly separable, local maximum separable clustering.

To define local maximum separability, several new definitions are required.
We call a point $x \in X$ a local maximum if $\epsilon_{N_p}(y) \geq \epsilon_{N_p}(x)$ for $y \in X$ such that $d(x, y) \leq \epsilon_{N_p}(x)$.
%For a cluster $c$, the set of local maximums is denoted by $\ell(c)$.
For a cluster $c \in C$, let $X_c^*$ denote the set of all points $x$ in $c$ such that $\epsilon_{N_p}(x) = \min_{y \in c} \epsilon_{N_p}(y)$.  In other words, $X_c^*$ denotes the set of all points $x$ in $c$ that have highest density among points in $c$.

For a given density parameter $N_p$, the relative separability of a point $x \in X$ to a point $y \in X$ is the value of $A$ such that $A \cdot \epsilon_{N_p}(x) = \epsilon(x, y)$, and is denoted by $A(x, y)$.  Similarly, the relative separability of a point $x \in X$ to a cluster $c$ is given by $\min_{y \in c} A(x, y),$ and is denoted by  $A(x, c).$
% \begin{Definition}
%     For a given density parameter $N_p$, the relative separability of a point $x \in X$ to a point $y \in X$ is the value of $A$ such that $A \cdot \epsilon_{N_p}(x) = \epsilon(x, y)$, and is denoted by $A(x, y)$.  Similarly, the relative separability of a point $x \in X$ to a cluster $c$ is given by $\min_{y \in c} A(x, y),$ and is denoted by  $A(x, c).$
%     % The relative separability of a cluster $c$ to another cluster $c'$ is given by $A(x_c^*, c'),$ and is denoted by $A(c, c')$.
% \end{Definition}
\begin{definition}
    For a given $N_p$ and $C$, we use $A^\ell(C)$ to denote the minimum $A \in \mathbb{R}$ such
    that 
    \begin{equation*}    \max_{y \in c(x): \; \epsilon_{N_p}(y) \leq \epsilon_{N_p}(x)} A(x, y) \leq A,
    \end{equation*}  
    for every local maximum $x \in X$ where there exists a $y \in c(x)$ such that $\epsilon_{N_p}(y) \leq \epsilon_{N_p}(x)$.
    %, and 
    % \begin{equation*}
    %    \min_{y \in c(x): \; \epsilon_{N_p}(y) = \epsilon_{N_p}(x)} A(x, y) \leq A
    % \end{equation*}
    % %\end{align*}
    % for every local maximum $x \in X$ where $\epsilon_{N_p}(x) = \min_{y \in c(x)} \epsilon_{N_p}(y)$ and there exists a $y \in c(x)$ such that $y \neq x$ and $\epsilon_{N_p}(y) = \epsilon_{N_p}(x)$.
    $C$ is called local maximum separable (LM-separable) if
    \begin{align*}
        A^\ell(C) < \min_{c, c' \in C} \min_{z \in X_c^*} A(z, c').
    \end{align*}
    In specific, if $C$ has only one local maximum per cluster, then $C$ is trivially LM-separable. 
\end{definition}
% \begin{Definition}
%     For a given density parameter $N_p,$ a clustering $C$ is called LM-separable if for every local maximum $x \in X$ where there exists a $y \in c(x)$ such that $\epsilon_{N_p}(y) < \epsilon_{N_p}(x)$, we have that 
%     \begin{align*}
%         \min_{y \in c(x): \; \epsilon_{N_p}(y) < \epsilon_{N_p}(x)} A(x, y) < \min_{c \in C} \min_{c' \in C, c' \neq c} A(x_c^*, c'),
%     \end{align*}  
%     and for every local maximum $x \in X$ such that $\epsilon_{N_p}(x) = \min_{y \in c(x)} \epsilon_{N_p}(y)$, we have that  
%     \begin{align*}
%         \min_{y \in c(x): \; \epsilon_{N_p}(y) = \epsilon_{N_p}(x)} A(x, y) < \min_{c \in C} \min_{c' \in C, c' \neq c} A(x_c^*, c'),
%     \end{align*}  
%     In specific, if $C$ has only one local maximum per cluster, then $C$ is trivially LM-separable.
%     We use $A^\ell(C)$ denote the minimum $A$ such that 
%     $\min_{y \in c(x): \; \epsilon_{N_p}(y) < \epsilon_{N_p}(x)} A(x, y) \leq A$
%     for every local maximum $x \in X$ that is not a cluster maximum, and     $\min_{y \in c(x): \; \epsilon_{N_p}(y) = \epsilon_{N_p}(x)} A(x, y) \leq A$ for every $x \in X$ that is a cluster maximum.     
%     % \begin{align*}
%     % & \min_{A: \; \exists y \in c(x) \cap c^*(x, A \cdot \epsilon_{N_p}(x)), \; \epsilon_{N_p}(y) \leq \epsilon_{N_p}(x) } A
%     % \\ & \leq \min_{A: \; \exists c, c' \in C, \; A \cdot \epsilon_{N_p}(x_c^*) = \epsilon(c, c')} A.
%     % \end{align*}
% \end{Definition}

%Note that $A^\ell(C)$ is also the the minimum $A$ such that $A(x, x_{c(x)}^*) \leq A$ for every local maximum $x\in X$. 

\begin{figure}
    \centering
    \includegraphics[width=1\linewidth]{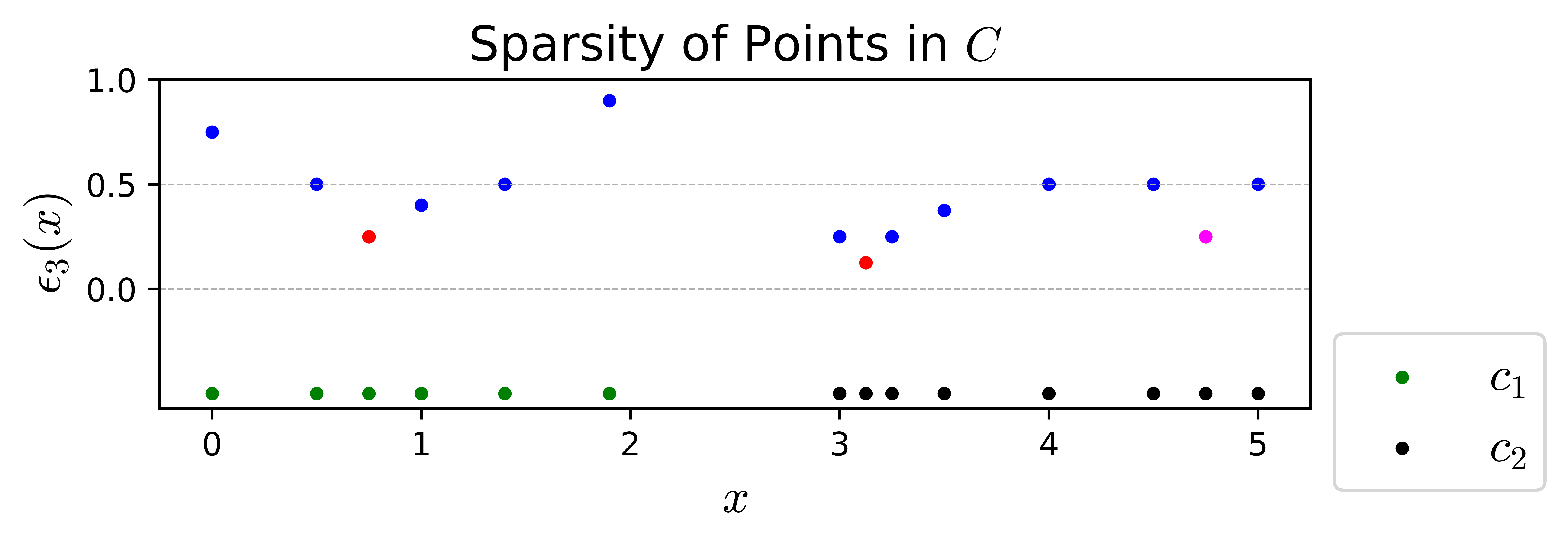}
    \caption{A plot showing the sparsity value $\epsilon_3(x)$ of each point $x \in X$ in a 14 point one-dimensional clustering $C = [c_1, c_2]$ that is weakly separable and LM-separable for $N_p = 3$, implying that recovery is guaranteed.  %The orange, red, and pink points correspond to local maxima.
    The red points correspond to points in $X_c^*$ for $c \in C$.  The pink point corresponds to the local maximum $x \in X$ such that $\max_{y \in c(x): \; \epsilon_{N_p}(y) \leq \epsilon_{N_p}(x)} A(x, y) = A^\ell(C)$ (i.e. the local maximum that looks the most separated from its cluster).   
    $C$ is weakly separable because $\epsilon^*(c_1) = 0.9$ and $\epsilon^*(c_2) = 0.5$, while $\min_{c, c' \in C} \epsilon(c, c') = 1.1$.
    $C$ is LM-separable because $A^\ell(C) = 2.0$ while $\min_{c, c' \in C} \min_{z \in X_c^*} A(z, c') = 4.4$. }
    \label{fig:sparsity_plot_simple}
\end{figure}

\begin{figure}
    \centering
    \includegraphics[width=1\linewidth]{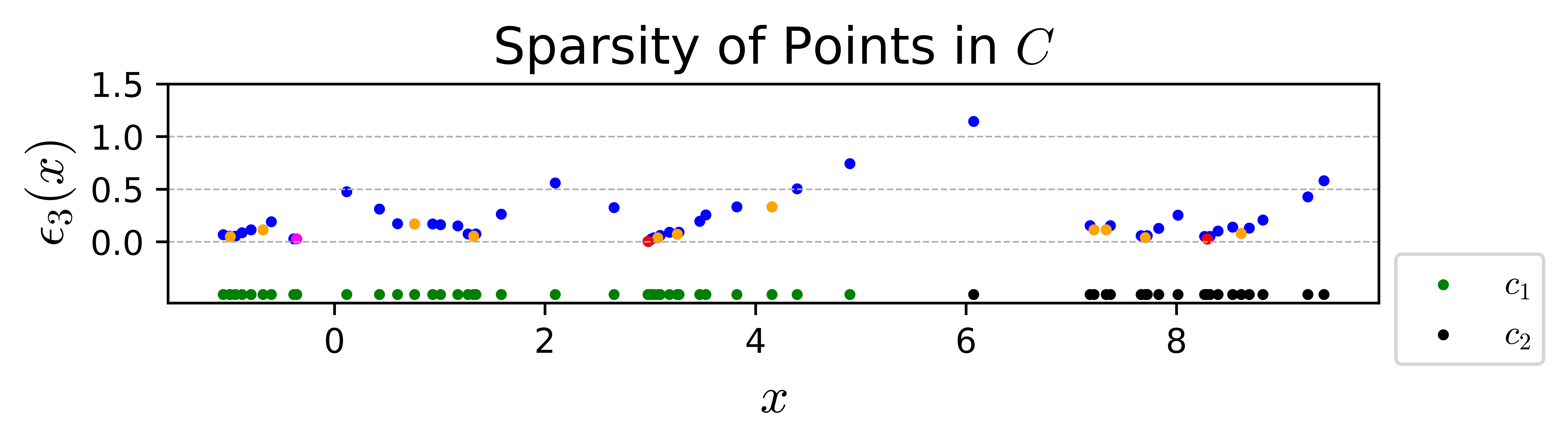}
    \caption{A plot showing the sparsity value $\epsilon_3(x)$ of each point $x \in X$ in a 60 point one-dimensional clustering $C = [c_1, c_2]$ that is weakly separable and LM-separable for $N_p = 3$, implying that recovery is guaranteed.  The orange, red, and pink points correspond to local maxima.  The red points correspond to points in $X_c^*$ for $c \in C$.  The pink point corresponds to the local maximum $x \in X$ such that $\max_{y \in c(x): \; \epsilon_{N_p}(y) \leq \epsilon_{N_p}(x)} A(x, y) = A^\ell(C)$ (i.e. the local maximum that looks the most separated from its cluster).   
    $C$ is weakly separable because $\epsilon^*(c_1) = 0.74$ and $\epsilon^*(c_2) = 1.14$, while $\min_{c, c' \in C} \epsilon(c, c') = 1.18$.
    $C$ is LM-separable because $A^\ell(C) = 20.49$ while $\min_{c, c' \in C} \min_{z \in X_c^*} A(z, c') = 44.27$. 
    %that has the highest value of  $\min_{y \in c(x): \; \epsilon_{N_p}(y) \leq \epsilon_{N_p}(x)} A(x, y)$ among all local maximums $x \in X$ where there exists a $y \in c(x)$ such that $\epsilon_{N_p}(y) \leq \epsilon_{N_p}(x)$.}
    }
    \label{fig:sparsity_plot}
\end{figure}

\begin{figure}%[htp]

\centering

    \begin{subfigure}{\includegraphics[width=0.45\linewidth]{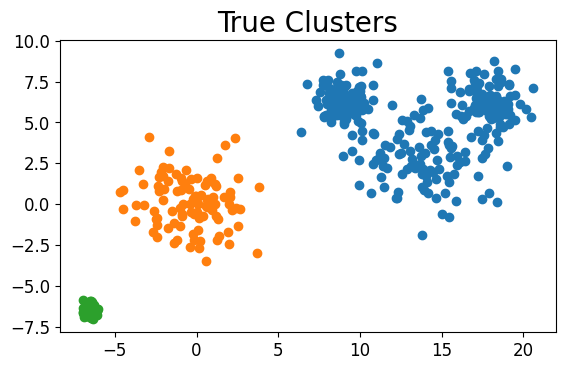}}
    \end{subfigure}
\hfill
    \begin{subfigure}{\includegraphics[width=0.45\linewidth]{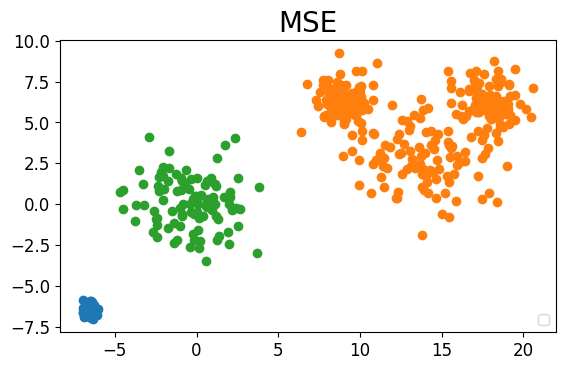}}
    \end{subfigure}
\hfill
    \begin{subfigure}{\includegraphics[width=0.45\linewidth]{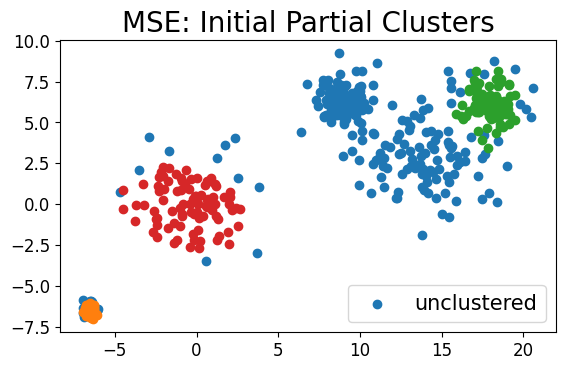}}
    \end{subfigure}
\hfill
    \begin{subfigure}{\includegraphics[width=0.45\linewidth]{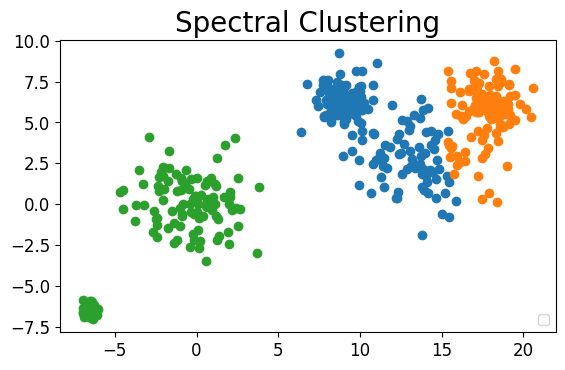}}
    \end{subfigure}

\caption{A clustering with 500 points that is weakly separable and LM-separable, but not strongly separable for $N_p = 5$, implying that recovery by our algorithm (``MSE'') is guaranteed. We compare against spectral clustering (scikit-learn) with the $4$-nearest neighbors affinity matrix because for $N_p = 5$, our algorithm uses the $4$ nearest neighbors of a point to calculate density.}
\vspace{-1.5em}
\label{fig:weak_not_strong}

\end{figure}

LM-separability specifies that for every local maximum $x \in X$ where there exists a $y \in c(x)$ whose density is at least as high as that of $x$, 
the relative separability of $x$ to any such $y$ is smaller than the relative separability of any highest density point in a cluster to another cluster.
This is a condition that should hold for a clustering to unambiguous because if there exists a local maximum $x$ where the relative separability to such a point $y \in c(x)$ is higher than the relative separability of a highest density point $z \in X_{c}^*$ for some $c \in C$ to another $c' \in C$ (and therefore to a local maximum in $c'$), then in a sense, $x$ looks more like it belongs to a separate cluster from $y$ than $z$ looks like it belongs to a separate cluster from $c'$. %and the correct $K$-clustering of $X$ is ambiguous.
Figures \ref{fig:sparsity_plot_simple}, \ref{fig:sparsity_plot}, and \ref{fig:weak_not_strong} illustrate LM-separability.
%Like strong separability, LM-separability is based on relative separability, and can thus be satisfied by clusterings with arbitrarily shaped clusters that have arbitrarily many relatively separated regions of high density and arbitrary variation in density among different clusters.
%arbitrary variation in density within clusters, and arbitrarily shaped clusters.  
%However,

% \begin{figure}
%     \centering
%     \includegraphics[width=1\linewidth]{Recovery of Unambiguous Clusters/images/sparsity_plot2.png}
%     \caption{A plot showing the sparsity value $\epsilon_3(x)$ of each point $x \in X$ in a 60 point one-dimensional clustering $C = [c_1, c_2]$ that is weakly separable and LM-separable for $N_p = 3$, implying that recovery is guaranteed.  The orange, red, and pink stars correspond to local maxima.  The red stars correspond to points in $X_c^*$ for $c \in C$.  The pink star corresponds to the local maximum $x \in X$ such that $\min_{y \in c(x): \; \epsilon_{N_p}(y) \leq \epsilon_{N_p}(x)} A(x, y) = A^\ell(C)$ (i.e. the local maximum that looks the most separated from its cluster).   
%     $C$ is weakly separable because $\epsilon^*(c_1) = 0.74$ and $\epsilon^*(c_2) = 1.14$, while $\min_{c, c' \in C} \epsilon(c, c') = 1.18$.
%     $C$ is LM-separable because $A^\ell(C) = 8.95$ while $\min_{c, c' \in C} \min_{z \in X_c^*} A(z, c') = 15.35$. 
%     %that has the highest value of  $\min_{y \in c(x): \; \epsilon_{N_p}(y) \leq \epsilon_{N_p}(x)} A(x, y)$ among all local maximums $x \in X$ where there exists a $y \in c(x)$ such that $\epsilon_{N_p}(y) \leq \epsilon_{N_p}(x)$.
%     }
%     \vspace{-1.5em}
%     \label{fig:sparsity_plot}
% \end{figure}

Note that LM-separability does not imply weak separability.  
Let $N_p = 2$, and consider the clustering $C = [\{7, 8, 10, 13, 21\}, \{17, 25, 27\}]$.  $C$ is clearly not weakly separable because the clusters overlap, but is LM-separable, as $7,8,25, 27$ are the only local maxima.

Our main result states that if $C$ is weakly separable and LM-separable for a given density parameter $N_p$, then it is the unique weakly separable, LM-separable clustering for $N_p$, and can be reconstructed efficiently.
\begin{theorem} \label{thm:LM_sep_main}
    If $C$ is weakly separable and LM-separable for $N_p$, then $C$ is the unique weakly separable, LM-separable clustering for $N_p$, and can be found in $O(|X|^3\log(|X|))$ time.
\end{theorem}
% Furthermore, if $C$ is LM-separated, then even if it is not weakly separated, we can still efficiently recover the $|C|$ cluster centers.  Note that this holds even if clusters partially overlap.  
% \begin{theorem} \label{thm:strong_sep_partial}
%     If $C$ is LM-separable for density parameter $N_p$, we can find the partial $|C|$-clustering given by $[\{x_c^*\} : c \in C]$ in $O(|X|^2\log(|X|))$ time.
% \end{theorem}

% Since strong separability implies LM-separability and weak separability, expanding an $\epsilon$-cut of the dendrogram is not sufficient for finding all weakly separable, LM-separable clusterings by Lemma \ref{lem:dbscan_insufficient}.

For a given $N_p$, it is in general not possible to recover the unique weakly separable, LM-separable clustering by simply finding an extension of the (partial) clustering given by the $\epsilon$-cut of the dendrogram that contains $K$ clusters (if there exists such an $\epsilon$)
as proved in Lemma \ref{lem:dbscan_insufficient}.  Since DBSCAN follows this approach, it is not sufficient for finding the unique weakly separable, LM-separable clustering.
%Intuitively, this is because a strongly separable clustering may have a high variation in density among different clusters that an $\epsilon$-cut cannot capture.   
\begin{lemma} \label{lem:dbscan_insufficient}
    There exists a weakly separable, LM-separable clustering for some $N_p$ that does not extend any (partial) clustering given by an $\epsilon$-cut of the dendrogram.
\end{lemma} 
\begin{IEEEproof}
    Recall that there can only be one $\epsilon$-cut of the dendrogram that gives a (partial) $K$-clustering.  Let $N_p = 3$, and suppose that 
    $C = [\{1, 3, 5, 7.02, 9.02, 11.02\},$ $\{17, 18, 19, 20\},$
        $\{22.01, 23.01, 24.01, 25.01\}].$
    % \begin{align}C = [\{1, 3, 5, 7.02, 9.02, 11.02\}, \{17, 18, 19, 20\},
    %     \{22.01, 23.01, 24.01, 25.01\}]. \nonumber
    % \end{align}
    % \begin{align*} C = [&\{1, 3, 5, 7.02, 9.02, 11.02\}, \{17, 18, 19, 20\}, 
    %     \\ & \{22.01, 23.01, 24.01, 25.01\}]. \end{align*}
    This is clearly weakly separable and LM-separable.  The only $\epsilon$-cut that gives a (partial) clustering with three clusters  
    is chosen by setting $\epsilon = 2.01$, and is given by 
    $C' = [\{3\}, \{9.02\}, $ 
         $\{17, 18, 19, 20, 22.01, 23.01, 24.01, 25.01\}].$
    % \begin{align*}
    %     C' = [&\{3\}, \{9.02\},
    %     \\ & \{17, 18, 19, 20, 22.01, 23.01, 24.01, 25.01\}].
    % \end{align*} 
    The only weakly separable clustering that is an extension of $C'$ is 
    $[\{1, 3, 5\},$ $\{7.02, 9.02, 11.02\},$ $\{17, 18, 19, 20, 22.01, 23.01, 24.01, 25.01\}].$ 
    % \begin{align} [\{1, 3, 5\}, \{7.02, 9.02, 11.02\}, \{17, 18, 19, 20, 22.01, 23.01, 24.01, 25.01\}]. \nonumber \end{align}
    % \begin{align*}
    %     [&\{1, 3, 5\}, \{7.02, 9.02, 11.02\},
    %     \\ & \{17, 18, 19, 20, 22.01, 23.01, 24.01, 25.01\}].
    % \end{align*} 
    %Thus, the partial $K$-clustering given by an $\epsilon$-cut of the dendrogram cannot be extended to a strongly separable clustering.
\end{IEEEproof}

%The counterexample in the proof of Lemma \ref{lem:dbscan_insufficient} contains three clusters, one of which is of lower density than the other two.  
In the clustering $C$ used to prove Lemma \ref{lem:dbscan_insufficient}, the distance separating the points $1, 3, 5$ from the points $7.02, 9.02, 11.02$ within the first cluster is $2.02$, which is larger than the distance of $2.01$ separating the clusters $\{17, 18, 19, 20\}$ and $\{22.01, 23.01, 24.01, 25.01\}$. However, $2.02$ is very similar to the distance of $2$ separating the other pairs of adjacent points in the first cluster, while $2.01$ is very large compared to the distance of $1$ separating the adjacent points in the second cluster.  $C$ is therefore a more natural clustering than $C'$ in a sense, but an $\epsilon$-cut is unable to capture $C$ because it only considers absolute distances when separating clusters, without taking cluster density into account.

In the appendix we discuss a stronger notion of separability called strong separability which implies weak separability and LM-separability. Lemma \ref{lem:dbscan_insufficient} holds for strong separability as well.

% \begin{corollary}
%     There exists a weakly separable, LM-separable clustering for some $N_p$ that does not extend any (partial) clustering given by an $\epsilon$-cut of the dendrogram.
% \end{corollary}
% \begin{proof}
%     This follows from Lemmas \ref{lem:strong-sep-implies_ell-sep} and \ref{lem:dbscan_insufficient}.
% \end{proof}

% We proceed to prove Theorems \ref{thm:LM_sep_main} 
%and \ref{thm:strong_sep_partial} 
% in the next section.

\section{Algorithm and Proof of Theorem \ref{thm:LM_sep_main}}

For density parameter $N_p$, we will prove that if $C$ is weakly separable and LM-separable, then Algorithm \ref{alg:cluster_alg} returns $C$, thus proving that $C$ is the unique weakly separable and LM-separable $|C|$-clustering for $N_p$.  We then prove that the algorithm can be implemented to run in $O(|X|^3 \log(|X|))$ time.

Algorithm \ref{alg:greedy_cluster} is an auxiliary algorithm used to define Algorithm \ref{alg:cluster_alg}.  Algorithm \ref{alg:greedy_cluster} works by outputting a partial $K$-clustering of $X$ that can then be postprocessed to form a $K$-clustering. For Algorithm \ref{alg:greedy_cluster}, $A$ governs the maximum variation in density within a partial cluster, $M$ governs the minimum size of a partial cluster, and $D$ governs the maximum variation in density among different partial clusters.  
We use $C_g(X, N_p, A, M, D)$ to denote the partial clustering of $X$ that is output by Algorithm \ref{alg:greedy_cluster} with parameters $N_p, A, M, D$. 

\begin{algorithm}
\caption{Minimal Seed Expansion (MSE)}
\label{alg:cluster_alg}   
\begin{algorithmic}[1]
    \REQUIRE{$X$, $N_p$, $M$, $D$, $K$}
    \ENSURE{$\hat{C}$}
    \STATE $C' \gets$ min-$A$ clustering of $X$ for $N_p, M, D, K$
    %\STATE $C' \! \gets$ output of Algorithm \ref{alg:greedy_cluster}  with input $X, N_p, M, D, K$ %(this is the min-$A$ partial clustering of $X$)
    \STATE $\hat{C} \gets$ output of Algorithm \ref{alg:greedy_expansion} with input $X, C', N_p$
\end{algorithmic}
\end{algorithm}

% \begin{algorithm}
% \caption{Minimal Seed Expansion}
% \label{alg:cluster_alg}   
%     \KwData{$X$, $N_p$, $M$, $D$, $K$}
%     \KwResult{$\hat{C}$}
%     $C' \gets$ min-$A$ partial clustering of $X$ for $N_p, M, D, K$\;
%     $\hat{C} \gets$ output of Algorithm \ref{alg:greedy_expansion} with parameters $X, C', N_p$\;
% \end{algorithm}
At a high level, Algorithm \ref{alg:greedy_cluster} forms a partial clustering of $X$ by greedily selecting the highest density unclustered point in $X$, and creating a maximal cluster centered at the point that satisfies the partial cluster constraints set by the parameters $A, M, D$.  Note that if there are multiple candidate points for $x^*$ in a greedy step of Algorithm \ref{alg:greedy_cluster}, the final output is not affected by which candidate is assigned to $x^*$.  Thus, for parameters $X, N_p, A, M, D$, the clustering $C_g(X, N_p, A, M, D)$ output by Algorithm \ref{alg:greedy_cluster} is unique.
% Note that the point $x^*$ chosen at each step of Algorithm \ref{alg:greedy_cluster} can be chosen arbitrarily from the set of maximum density unclustered points.

\begin{definition}
    The min-$A$ clustering of $X$ for $N_p, M, D, K$ is given by $C_g(X, N_p, A_{min}, M, D)$ where     \begin{equation*}
        A_{min} = \min\{A: |C_g(X, N_p, A, M, D)| = K \}.
        % A_{min} =  \min_{A \text{ such that } |C_g(X, N_p, A, M, D)| = K } A.
    \end{equation*}
    For parameters $N_p, M, D, K$, the min-$A$ clustering is denoted by \[C_m(X, N_p, M, D, K).\]  
    $A_{min} \geq 1$ since $c^*(x, \epsilon) = \{\}$ if $\epsilon < \epsilon_{N_p}(x)$.
\end{definition}

Algorithm \ref{alg:cluster_alg} accepts $N_p, M, D$ along with a number of clusters $K$ as input, and begins by finding the min-$A$ partial clustering for $N_p, M, D, K$.
The min-$A$ partial clustering is then passed to Algorithm \ref{alg:greedy_expansion} to produce the final output clustering.

% \begin{Definition}
%     For a given density parameter $N_p$ and number of clusters $K,$ a min-$A$ partial clustering is defined as any clustering $C_g(X, N_p, A^*, M, D)$ where 
%     \begin{align}
%         A^* =  \min_{A \text{ such that } |C_g(X, N_p, A, M, D)| = K } A,
%     \end{align}
%     and is denoted by $C_m(X, N_p, M, D, K)$.  
%     We use $A_{min}$ to denote $A^*$.
% \end{Definition}

% Clearly $A_{min} \geq 1$ for any $X, N_p, K$ since $c^*(x, \epsilon) = \{\}$ if $\epsilon < \epsilon_{N_p}(x)$.

\begin{algorithm}
\caption{Greedy algorithm to find partial clusters}
\label{alg:greedy_cluster}
\begin{algorithmic}
    \REQUIRE{$X$, $N_p$, $A$, $M$, $D$}
    \ENSURE{$C'$}
    \STATE $C' \gets \emptyset$
    \STATE $\text{MinExtracted} \gets \infty$ \STATE $\text{Tried} \gets \emptyset$
    % \STATE Compute $\epsilon_{N_p}(x)$ for all $x \in X$
    % \STATE $C' \gets \emptyset
    % \STATE $\text{MinExtracted} \gets \infty$
    % \STATE $\text{Tried} \gets \emptyset$
    \WHILE{ $X \setminus \text{Tried} \neq \emptyset$}
        \STATE $x^* \gets \arg \min_{x \in X \setminus \text{Tried}} \epsilon_{N_p}(x)$
        \IF{$\epsilon_{N_p}(x^*) > D \cdot \text{MinExtracted}$}
            \STATE break loop
        \ENDIF
        \STATE $c' \gets c^*(x^*, A\cdot \epsilon_{N_p}(x^*))$
        \IF{$|c'| \geq M$ and $c' \cap c = \emptyset \;\;  \forall c \in C'$ }
            \STATE add $c'$ to $C'$\IF{$\text{MinExtracted} = \infty$}
            \STATE $\text{MinExtracted} \gets \epsilon_{N_p}(x^*)$
            \ENDIF

            \STATE $X \gets X \setminus c'$
        \ELSE
            \STATE $\text{Tried} \gets \text{Tried} \cup \{x^*\}$
        \ENDIF
    \ENDWHILE
\end{algorithmic}
\end{algorithm}

%\vspace{-1.5em}
\begin{algorithm}
\caption{Greedy algorithm to expand partial clusters}
\label{alg:greedy_expansion}   
\begin{algorithmic}
    \REQUIRE{$X$, $C'$}
    \ENSURE{$\hat{C}$}
    \STATE $\hat{C} \gets C'$ 
    \STATE $Y \gets X \setminus (\cup_{c \in C'} c)$
    % \STATE $Y \gets X \setminus (\cup_{c \in C'})$
    \WHILE{$\cup_{c \in \hat{C}} c \neq X $} 
        \STATE $(x^*, c^*) \gets \arg \min_{x \in Y, c \in \hat{C}} \epsilon(x, c)$
        \STATE $c^* \gets c^* \cup \{x^*\}$ 
        \STATE $Y \gets Y \setminus \{x^*\}$
        % \STATE $Y \gets Y \setminus x^*$
    \ENDWHILE
\end{algorithmic}
\end{algorithm}

% \begin{algorithm}
% \caption{Greedy algorithm to find partial clusters}
% \label{alg:greedy_cluster}
%     \KwData{$X$, $N_p$, $A$, $M$, $D$}
%     \KwResult{$C'$}
%     $C' \gets \emptyset$\;
%     % \STATE Compute $\epsilon_{N_p}(x)$ for all $x \in X$\;
%     $\text{MinExtracted} \gets \infty$\;
%     $\text{Tried} \gets \emptyset$\;
%     \While{ $X \setminus \text{Tried} \neq \emptyset$}
%     {
%         $x^* \gets \arg \min_{x \in X \setminus \text{Tried}} \epsilon_{N_p}(x)$\;
%         \If{$\epsilon_{N_p}(x^*) > D \cdot \text{MinExtracted}$}
%         {
%             break loop\;
%         }
%         $c' \gets c^*(x^*, A\cdot \epsilon_{N_p}(x^*))$\;
%         \If{$|c'| \geq M$ and $c' \cap c = \emptyset \;\;  \forall c \in C'$ }
%         {    
%             add $c'$ to $C'$\;
%             \If{$\text{MinExtracted} = \infty$}
%             {
%             $\text{MinExtracted} \gets \epsilon_{N_p}(x^*)$\;
%             }
%             $X \gets X \setminus c'$\;
%         }          
%         \Else{
%             $\text{Tried} \gets \text{Tried} \cup \{x^*\}$\;
%         }
%     }
% \end{algorithm}

% \begin{algorithm}
% \caption{Greedy algorithm to expand partial clusters}
% \label{alg:greedy_expansion}   
%     \KwData{$X$, $C'$}
%     \KwResult{$\hat{C}$}
%     $\hat{C} \gets C'$\;
%     $Y \gets X \setminus (\cup_{c \in C'} c)$\;
%     \While{$\cup_{c \in \hat{C}} c \neq X$} 
%     {
%         $(x^*, c^*) \gets \arg \min_{x \in Y, c \in \hat{C}} \epsilon(x, c)$\;
%         $c^* \gets c^* \cup \{x^*\}$\;
%         $Y \gets Y \setminus \{x^*\}$\;
%     }
% \end{algorithm}

\begin{lemma} \label{lem:ell-sep-lemm-helper}
    For a given $N_p,$ if $C$ is LM-separable, then \[|C_g(X, N_p, A^\ell(C), 1, \infty)| = K.\]
\end{lemma}
\begin{IEEEproof}
    We will prove that all partial clusters output by Algorithm \ref{alg:greedy_cluster}  are subsets of distinct clusters in $C$.  Consider the $i$th partial cluster output by Algorithm \ref{alg:greedy_cluster}.  The highest density point $x$ that the algorithm uses to build the $i$th cluster is clearly a local maximum, and can either be from a previously partially reconstructed cluster in $C$, or can be from a new cluster in $C$.  If $x$ is from a previously partially reconstructed cluster $c' \in C$, then the $i$th partial cluster must include some $z \in X_{c'}^*$ from a previously output partial cluster by the definitions of LM-separability and $A^\ell(C)$, and the fact that $\epsilon_{N_p}(x) \geq \epsilon_{N_p}(z).$  This is a contradiction.  This, along with the fact that $M = 1 \leq N_p$, guarantees that $x$ must be from a new cluster $c'' \in C$ and $x \in X_{c''}^*$.  The $i$th partial cluster $c^*(x, A^\ell(C) \cdot \epsilon_{N_p}(x))$ must only include points from $c''$ by the definition of LM-separability and $A^\ell(C)$. 
    Because only a partial cluster of a new cluster in $C$ can be output in each iteration of the algorithm, a partial $K$-clustering is output.
    %Because only a new cluster center can be added to the partial clustering in each iteration of the algorithm, a partial $K$-clustering is output.
\end{IEEEproof}
By Lemma \ref{lem:ell-sep-lemm-helper},  $A_{min}$ is no larger than $A^\ell(C)$.

\begin{lemma} \label{lem:ell-sep-lemma}
    For a given $N_p,$ if $C$ is weakly separable and LM-separable, the min-$A$ clustering $C_m(X, N_p, 1, \infty, |C|)$ is extendable to $C$.
\end{lemma}
\begin{IEEEproof}   
    %We will prove this by induction on $K$.  Suppose the property holds for all clusterings with $K-1$ clusters.   
    %We will prove that the partial clustering output by Algorithm \ref{alg:greedy_cluster} is extendable to $C$.
    Consider the $i$th partial cluster output by Algorithm \ref{alg:greedy_cluster}.  The highest density point $x$ used by algorithm to build the $i$th cluster must be a local maximum, and can either be from a previously partially reconstructed cluster in $C$, or can be from a new cluster in $C$. 
    
    If $x$ is from a previously partially reconstructed cluster $c \in C$, the $i$th partial cluster does not include any points from other true clusters that are not yet part of a previously output partial cluster.  This is because if it did include a point from such a cluster $c' \in C$, it would include a point $z \in X_{c'}^*$ by weak separability of $C$, and since $\epsilon_{N_p}(z) \geq \epsilon_{N_p}(x)$, it would imply that $A_{min} \cdot \epsilon_{N_p}(z) \geq \epsilon(c', c)$ which violates LM-separability of $C$ since $A_{min} \leq A^\ell(C)$.  

    If the highest density point $x$ is from a new true cluster $c \in C$, by LM-separability of $C$, only points from $c$ are added to the partial cluster since $A_{min} \leq A^\ell(C)$.

    %Thus, the algorithm outputs a set of at least $K$ non-overlapping partial clusters such that for each 
    
    Therefore, the algorithm outputs non-overlapping partial clusters such that each partial cluster chosen does not include any points from true clusters that are not yet part of a previously output partial cluster.
    Observe that, if no points in a cluster $c \in C$ have been added to previously output partial clusters by the time $w \in X_c^*$ is selected as the highest density unclustered point by the algorithm, then a partial cluster centered at $w$ that is a non-empty subset of $c$ will be output by the algorithm in that step since $M = 1 \leq N_p$, $A_{min} \leq A^\ell(C)$ and $C$ is LM-separable. 
    Therefore, we obtain a partial clustering that contains a partial cluster centered at $w \in X_c^*$ that is a  non-empty subset of $c$ for each $c \in C$.
    %Therefore, we obtain a partial clustering that contains a partial cluster of $c$ centered at $w \in X_c^*$ for each $c \in C$.
    %If this partial clustering  contains more than $K$ clusters, then this contradicts the fact that it is a min-$A$ partial clustering for $K$. 
    This is a min-$A$ partial clustering for $K$, so it is  %the algorithm will output 
    a partial $K$-clustering that contains a partial cluster centered at some $w \in X_c^*$ that is a non-empty subset of $c$ for each $c \in C$.
    %A partial cluster centered at $w \in X_{c}^*$ for $c \in C$ only contain points from $c$ by LM-separability of $C,$ and the fact that $A_{min} \leq A^\ell(C)$.
    Thus, the algorithm outputs a partial $K$-clustering that is extendable to $C$.  
\end{IEEEproof}

\begin{lemma} \label{lem:well_sep_recovery}
    For a given $N_p$, if $C$ is weakly separable, and Algorithm \ref{alg:greedy_expansion} is initialized with a partial clustering $\hat{C}$ that is extendable to $C$, then Algorithm \ref{alg:greedy_expansion} recovers $C$.
\end{lemma}

\begin{IEEEproof} 
    Suppose that at some step in Algorithm \ref{alg:greedy_expansion}, a point $x$ is added to a cluster $c' \neq c(x)$ with some $\epsilon$.  This implies that at this step, there is no point $y \in c(x)$ that is not yet in $\hat{C}$ that is $\epsilon$-connected to another point $z \in c(x)$ that is already in $\hat{C}$.  
    This is a contradiction to the weak separability of $C$.  
    %In specific, there must be a point that is $\epsilon$-connected to a point in $c(x)$ that is already in $\hat{C}$ because $c(x)$ is $\epsilon$-connected for some $\epsilon < \min_{c'\in C, \; c' \neq c }\epsilon(c, c')$.
\end{IEEEproof}

Lemma \ref{lem:ell-sep-lemma} and Lemma \ref{lem:well_sep_recovery} imply that if $C$ is weakly separable and LM-separable, then it is the unique weakly separable, LM-separable clustering for $X$ and $N_p$.
As a consequence of Lemmas \ref{lem:min-A_runtiime} and \ref{lem:expansion_runtime}, Algorithm \ref{alg:cluster_alg} can be implemented in $O(|X|^3 \log(|X|))$ time.  

\begin{lemma} \label{lem:min-A_runtiime}
    For a given $X, N_p, M, D,$ the min-$A$ clustering can be found in $O(|X|^3 \log(|X|))$ time.
\end{lemma}
\begin{IEEEproof}
    %Notice that 
    We implement the greedy step of Algorithm \ref{alg:greedy_cluster} to have a deterministic rule for deciding whether to choose $x^*$ to be $x$ or $y$ if  $\epsilon_{N_p}(x) = \epsilon_{N_p}(y)$ (such as taking the point that comes first in the dataset).
    
    For a given $X, N_p, M, D,$ the number of clusters in $C_g(X, N_p, A, M, D)$ monotonically decreases as $A$ increases because of the following property.  Consider some $a, a' \geq 1$ such that $a > a'$. At the end of the $i$th iteration of Algorithm \ref{alg:greedy_cluster},
    the set of remaining candidates for future $x^*$ where $c^*(x^*, A\cdot \epsilon_{N_p}(x^*))$ will not intersect with a previously output cluster in the case when $A = a'$ is a superset of the set of such candidates when $A = a$.
    %if $A = a'$, there must be at least as many remaining candidates for future $x^*$ where $c^*(x^*, A\cdot \epsilon_{N_p}(x^*))$ will not intersection with a previously output cluster compared to the case where $A=a.$
    
    We will prove this by induction. Suppose this property holds for the $(i-1)$th iteration of Algorithm \ref{alg:greedy_cluster}, and consider the $i$th iteration. The point $x_{a'}$ picked to be $x^*$ by the algorithm if $A = a'$ may or may not already be included in a cluster previously output by the algorithm with $A = a$.  If not, then $x_{a'}$ will be chosen as $x^*$ for the algorithm with $A = a$ by the inductive hypothesis, so clearly at the end of the $i$th step, the property still holds since $a > a'$.
    
    %the number of suitable candidates remaining for the algorithm with $A = a$ is no larger than the number of suitable candidates remaining for the algorithm with $A = a'$.
    
    If on the other hand, $x_{a'}$ is included in a cluster previously output by the algorithm with $A = a$, then all points $x$ that are 
    $(A \cdot \epsilon_{N_p}(x))$-connected to $x_{a'}$ already cannot be candidates at the beginning of the $i$th step for the algorithm with $A = a$.  Therefore, in this case, at the end of the $i$th step,  the property still holds.
    %the number of suitable candidates remaining for the algorithm with $A = a$ is no larger than the number of suitable candidates remaining for the algorithm with $A = a'$.

    % This is because any local maximum $x$ that was included in a previously output cluster is such that any point $y$ that is $(A \cdot \epsilon_{N_p}(x))$-connected to $x$ cannot be picked as $x^*$ in a future cluster because $x$ is $(A \cdot \epsilon_{N_p}(y))$-connected to $y$ since $\epsilon_{N_p}(x) \leq \epsilon_{N_p}(y)$.
    
    Due to the monotonicity property we have just proved, if we have a set $S$ that includes all $A$ values that lead to distinct clusterings $C_g(X, N_p, A, M, D)$, then we can use binary search on $S$ to find the min-$A$ clustering for $X,$ $N_p,$ $M,$ $D.$  Algorithm \ref{alg:greedy_cluster} runs in $O(|X|^2)$ time.  Therefore, if we have such a set $S$ and the sorted list of its elements, this binary search approach to find the min-$A$ clustering runs in $O(\log(|S|)\cdot|X|^2)$ time.  Sorting the elements in $S$ for use in binary search runs in $O(|S| \log(|S|))$ time.  

    Consider a point $x^*$ selected in a greedy step of  Algorithm \ref{alg:greedy_cluster}.  One set that includes all $A$ values that could lead to different clusterings is given by $T_{x^*} = \{d(y, z) / \epsilon_{N_p}(x^*) : \; y, z \in X\}$ i.e. the set of all distances between points in $X$ divided by $\epsilon_{N_p}(x^*)$.
    Thus, $S = \cup_{x \in X} T_x$ includes every possible $A$ value that could lead to a different clustering.  We have that $|S| = O(|X|^3)$.  Thus, the binary search to find the min-$A$ clustering runs in $O(|X|^2 \log(|X|))$ time.  Constructing $S$ runs in $O(|X|^3)$ time, and sorting the values of $S$ for binary search runs in $O(|X|^3 \log(|X|))$ time. 
    The total runtime of the approach is therefore $O(|X|^3 \log(|X|))$.
\end{IEEEproof}

\begin{lemma} \label{lem:expansion_runtime}
    For a given $X, N_p, \hat{C}$, Algorithm \ref{alg:greedy_expansion} can be implemented to run in $O(|X|^2)$ time.
\end{lemma}
\begin{IEEEproof}
    Observe that for each greedy step of Algorithm \ref{alg:greedy_expansion}, any point $x \in X \setminus (\cup_{c \in \hat{C}} c)$ and cluster $c \in \hat{C}$ that minimize the quantity  $\epsilon^!(x, c) = \min_{y \in c} \max(d(x, y), \epsilon_{N_p}(x), \epsilon_{N_p}(y))$ can be selected as $(x^*, c^*)$.
    To see this, consider the set $S$ of unclustered points that are closest to some cluster in terms of $\epsilon$-distance.  Denote this minimum $\epsilon$-distance by $\epsilon'$.  Clearly, $S$ must include an unclustered point that contains a clustered point that it is $\epsilon'$-connected to in its $\epsilon'$-ball.
    Thus, at each greedy step of Algorithm \ref{alg:greedy_expansion}, we can pick $(x, c)$ that minimizes $\epsilon^!(x, y)$.

    To initialize the algorithm, for each unclustered point $x \in X \setminus (\cup_{c \in \hat{C}} c)$, and each clustered point $y$, we compute  $ \max(d(x, y), \epsilon_{N_p}(x), \epsilon_{N_p}(y))$ which then allows us to compute $\epsilon^!(x) = \min_{c \in \hat{C}}\epsilon^!(x, c)$ and $c^!(x) = \arg \min_{c \in \hat{C}}\epsilon^!(x, c)$ for every $x \in  X \setminus (\cup_{c \in \hat{C}} c)$. 
    %and $c \in \hat{C}$.  
    This can be done in $O(|X|^2)$ time.

    For each greedy step of Algorithm \ref{alg:greedy_expansion}, we simply set $(x^*, c^*)$ equal to the $(x, c)$  that minimizes $\arg \min_{c \in \hat{C}}\epsilon^!(x, c)$  in $O(|X|)$ time by checking $(\epsilon^!(x)$, $c^!(x))$ for all $x \in  X \setminus (\cup_{c \in \hat{C}} c)$.
    %$\epsilon_1(x, y)$. 
    After assigning $x^*$ to $c^*$, we update $\epsilon^!(x)$ %$\epsilon_1(x, c^*)$ 
    for every remaining unclustered point $x$ by setting 
    \begin{align}
    \epsilon^!(x) \gets \min(\epsilon^!(x), \; \max(d(x, x^*), \; \epsilon_{N_p}(x), \; \epsilon_{N_p}(x^*))), \nonumber
    \end{align}  and setting $c^!(x) \gets c^*$ if the value of $\epsilon^!(x)$ is changed.  
    %$\epsilon_1(x, c^*) \gets \min(\epsilon_1(x, c^*), \; \max(d(x, x^*), \; \epsilon_{N_p}(x), \; \epsilon_{N_p}(x^*)))$.  
    Each of these greedy steps takes $O(|X|)$ 
    %$O(|\hat{C}|\cdot|X|)$ 
    time and there are at most $O(|X|)$ greedy steps.  This stage of the algorithm therefore runs in $O(|X|^2)$ time.
    %$|\hat{C}|\cdot |X|^2$   
    % sort the other points in $X$ according to distance from $x$.  Performing this sorting for every $x \in X$ runs in $O(|X|^2 \log(|X|))$ time.
\end{IEEEproof}

\section{Experiments}

We compare a modified version of Algorithm \ref{alg:cluster_alg} for handling overlapping clusters to widely used algorithms for non-convex cluster recovery on a range of datasets. 
This modified version uses a variation of Algorithm \ref{alg:greedy_cluster} where each time a cluster is output, the points in the cluster are removed from $X$. 
This makes the check of whether a new cluster intersects with a previously output cluster inapplicable since only unclustered points remain in $X$ at the beginning of each greedy step. %of We use this approach because it appears to work well on datasets that have substantial cluster overlap. %, as in the case of many real datasets.
To increase speed, instead of finding the minimum $A$ that outputs a $K$-clustering, the implementation approximates this value of $A$ by progressively adjusting $A$ until a $K$-clustering is output by the modified version of Algorithm \ref{alg:greedy_cluster}.  To improve performance slightly, we use $N_p=2$ in Algorithm \ref{alg:greedy_expansion}, regardless of the $N_p$ used for finding the initial partial clusters.  
%This resulted in a small increase in performance in some cases.
We implemented this modified version of Algorithm \ref{alg:cluster_alg} in Python, and refer to it as ``MSE'' in this section.
%We refer to our algorithm as ``MSE,'' which stands for minimal seed expansion. We implemented the in Python.
%We implement this algorithm in Python, and refer to it ``MSE.'' 
Treating the dimensionality of the data as a constant, this implementation runs in $O(t  |X|^2)$ time where $t$ is the number of $A$ values tried.
The code for this implementation is available at: \url{https://github.com/kmazooji/Minimal-Seed-Expansion}.

We compare our algorithm to $K$-means, spectral clustering, HDBSCAN, OPTICS, and SpectACl \cite{hess2019spectacl}.
Treating the dimensionality of the data as a constant and letting $t$ be the number of iterations of $K$-means, $K$-means runs in $O(t K |X|)$ time, spectral clustering with k-nearest neighbors affinity runs in $O(K |X|^2 + t K |X|)$ time, HDBSCAN runs in $O(|X|^2)$ time, OPTICS runs in $O(|X|^2)$ time, and SpectACl runs in $O(d |X|^2 + t K |X|)$ time where $d$ is the embedding dimension. 

We use the spectral clustering and $K$-means implementations from scikit-learn, where the number of clusters $K$ is specified by the user.  The affinity matrix in spectral clustering is formed using each point's $K_n$ nearest neighbors, where $K_n$ is by the user.  The implementation of $K$-means uses the ``greedy $K$-means++'' algorithm \cite{ostrovsky2013effectiveness, vassilvitskii2007kmeans}.  %HDBSCAN, and OPTICS are very popular density based clustering algorithms that outputs whatever number of clusters it deems appropriate. 
We use the HDBSCAN implementation from the Python HDBSCAN clustering library, and we run it using the default cluster selection criteria named Excess of Mass (eom).  
The default setting for HDBSCAN sets $N_p$ equal to $M+1$ where $M$ is the minimum possible cluster size set by the user.
%In our experiments, 
We refer to this default version as ``HDBSCAN.''  
%In our experiments, 
We refer to the version of HDBSCAN where $N_p$ and $M+1$ are not tied as ``HDBSCAN (2).''
We use the OPTICS implementation from scikit-learn, where $N_p$, the minimum cluster size $M$, and the cluster selection parameter $X_i$ are set by the user.  Neither HDBSCAN nor OPTICS uses the number of clusters $K$.
SpectACl uses knowledge of $K$ and accepts a parameter $\epsilon$.  We use the implementation available on the TU Dortmund website from the authors: \url{https://sfb876.tu-dortmund.de/spectacl/index.html}. %, and use the default setting where $\epsilon$ is set by the user.  

\begin{figure}%[htp]
\centering

    \begin{subfigure}{\includegraphics[width=0.46\linewidth]{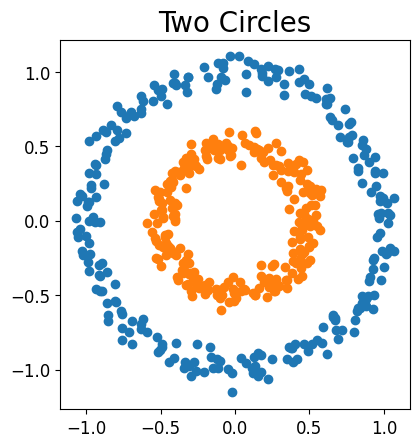}}
    \end{subfigure}
\hfill
    \begin{subfigure}{\includegraphics[width=0.64\linewidth]{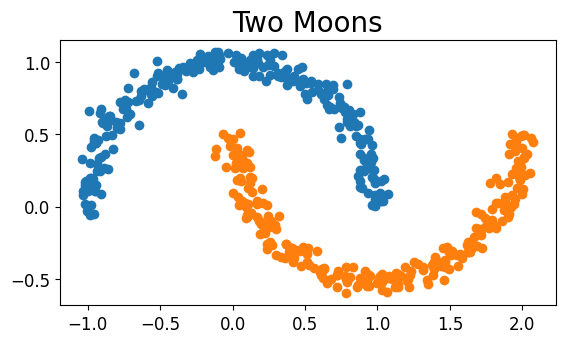}}
    \end{subfigure}
\hfill
    \begin{subfigure}{\includegraphics[width=0.54\linewidth]{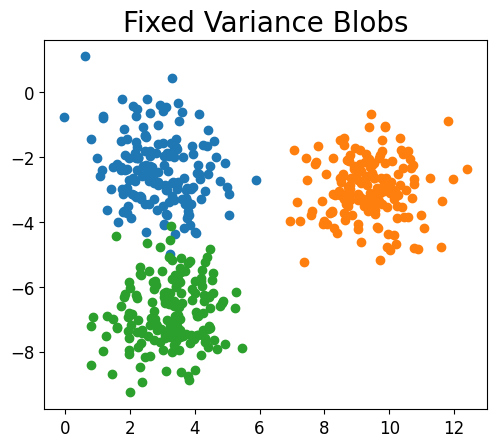}}
    \end{subfigure}
\hfill
    \begin{subfigure}{\includegraphics[width=0.5\linewidth]{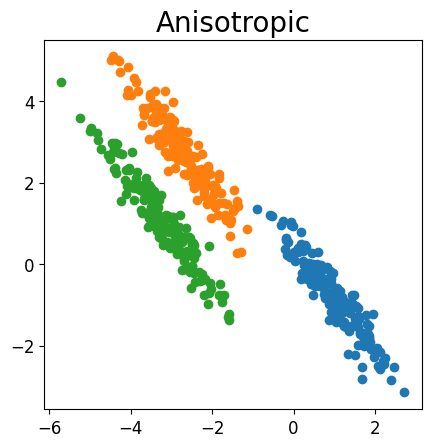}}
    \end{subfigure}
\hfill
    \begin{subfigure}{\includegraphics[width=0.5\linewidth]{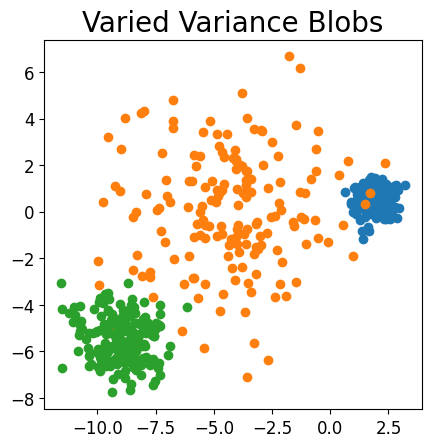}}
    \end{subfigure}

\caption{Artificial datasets used in Table \ref{table:simulation-results-sweep} .}

\label{fig:artificial_datasets}

\end{figure}

The Adjusted Rand Index (ARI) and Normalized Mutual Information (NMI) are the most common measures of similarity between an estimated clustering and a ground truth clustering.  For each algorithm, we report both measures for each dataset tested.    
The algorithms' performance is compared on the real world benchmark datasets whose properties are in Table \ref{table:benchmark-datasets}.  All of these datasets are available on the UCI server \cite{Dua:2019}. %  Iris, Wine, Seeds, Glass, Cancer, Digits, Letters and MNIST, which are all available on the UCI server \cite{Dua:2019}.
% \iscomment{change the name from ``Breast'' to ``Cancer''}
% \kmcomment{Done}
``Cancer'' refers to the Breast Cancer Wisconsin (Diagnostic) dataset.  ``Digits'' refers to the test set of the Optical Recognition of Handwritten Digits dataset.
``Letters'' refers to the test set of the Letter Recognition dataset.  ``MNIST'' refers to the test set of the MNIST dataset.  
%The properties of the datasets are given in Table \ref{table:benchmark-datasets}.  
For the MNIST dataset, we used t-SNE to reduce the dimensionality to two 
%\iscomment{many people consider ``wrong'' to use t-SNE and then clustering}
\cite{van2008visualizing}. We also compare the same algorithms for 5 artificial datasets given in Figure \ref{fig:artificial_datasets}  %created using scikit-learn 
which were used in the example titled ``Comparing different clustering algorithms on toy datasets'' on the scikit-learn website \cite{scikitlearn_toy_datasets}.  The properties of the datasets are given in Table \ref{table:artificial-datasets}.

\begin{table} 
\begin{center}
\caption{ \label{table:benchmark-datasets} benchmark datasets and their properties}
\begin{tabular}{ |c|c|c|c|c| } 
 \hline
dataset & \# points  & \# features & \# clusters & min. cluster size\\
\hline
 Iris & 150 & 4 & 3 & 50 \\
 \hline
 Wine & 178 & 13 & 3 & 48 \\
 \hline
 Seeds & 210 & 7 & 3 & 70 \\
 \hline
 Glass & 214 & 9 & 7 & 9\\
 \hline
 Cancer & 556 & 30 & 2 & 212 \\
 \hline
 Digits & 1,797 & 64 & 10 & 174 \\
 \hline
 Letter & 4,000 & 16 & 26 & 132 \\
 \hline
 MNIST & 10,000 & 784 & 10 & 892 \\
 \hline
\end{tabular}
\end{center}
\end{table}

The first set of results for benchmark datasets are reported in Table \ref{table:benchmark-results-sweep}. %, and results for artificial datasets are in Table \ref{table:simulation-results-sweep}.   
%For our algorithm, we set $N_p = 3$, pick $M$ to be a lower bound on the minimum cluster size.  
In these experiments, for MSE, we did not optimize $N_p$, $M$ and $D$ exhaustively. Instead, we set $N_p=3$, picked $M$ smaller than the true minimum cluster size for each dataset to give a competitive ARI, and used the $D$ value from the set $\{1.5, 2, 20\}$ that gave the best ARI.
%we did not optimize $N_p$, $M$ and $D$ in a meaningful way.  For all experiments we set $N_p=3$, picked $M$ values that seemed reasonable, and used the $D$ value from the set $\{1.5, 2, 10\}$ that gave the best results.

\begin{table} 
\begin{center}
\caption{ \label{table:benchmark-results-sweep} 
$|\hat{C}|$ is the number of clusters output by the algorithm.  For HDBSCAN and OPTICS, the set of noise points counts as a cluster.
%For the MNIST dataset, we used TSNE to reduce the dimensionality to 2. 
 }
%\resizebox{1\columnwidth}{!}{
\begin{tabular}{ |c|c|c|c|c| } 
 \hline
dataset & algorithm  & ARI & NMI & $|\hat{C}|$ \\
 \hline \multirow{5}{4em}{Iris}
 & MSE & {\bf 0.886} & {\bf 0.871} & 3\\ 
 & Spectral & 0.835 & 0.833 & 3\\
 & $K$-means & 0.716 & 0.742 & 3 \\
 & HDBSCAN & 0.568 & 0.734  & 2 \\
 & HDBSCAN (2) & 0.568 & 0.734  & 2 \\
 % & OPTICS & 0.732 & 0.753 & 4 \\
 & OPTICS & 0.732 & 0.753 & 4 \\
 & SpectACl & 0.653 & 0.682 & 3 \\
 \hline \multirow{5}{4em}{Wine} 
 & MSE & {\bf 0.439} & 0.430  & 3\\ 
 & Spectral & 0.401 & 0.395  & 3 \\ 
 & $K$-means & 0.371 & 0.429 & 3 \\
 & HDBSCAN & 0.292 & 0.379  & 2\\ 
 & HDBSCAN (2) & 0.291 & 0.403  & 3\\ 
 % & OPTICS & 0.327 & 0.428 & 2 \\ 
 & OPTICS & 0.418 & 0.405 & 3 \\
 & SpectACl & 0.427 & {\bf 0.462} & 3 \\
 \hline \multirow{5}{4em}{Seeds}
 & MSE & {\bf 0.725} & 0.682  & 3 \\ 
 & Spectral & 0.657 & 0.660  & 3 \\ 
 & $K$-means & 0.717 & {\bf 0.695} & 3 \\
 & HDBSCAN & 0.336 & 0.468  & 5 \\
 & HDBSCAN (2) & 0.409 & 0.469  & 3 \\
 % & OPTICS & 0.547 & 0.594 & 3 \\
 & OPTICS & 0.551 & 0.573 & 3 \\
 & SpectACl & 0.631 & 0.610 & 3 \\
 \hline \multirow{5}{4em}{Glass}
 % & MSE & 0.191 & 0.374  & 7 \\ %This is for M = 8 
 & MSE & 0.232 & 0.379  & 7 \\ %This is for M = 3
 & Spectral & 0.202 & 0.367  & 7  \\
 & $K$-means & 0.216 & 0.388 & 7 \\
 & HDBSCAN & 0.277 & 0.446 & 6 \\
 %& HDBSCAN (2) & 0.222 & 0.395  & 6 \\ M = 8
 & HDBSCAN (2) & 0.216 & 0.381  & 8 \\
 % & OPTICS & 0.264 & 0.440 & 4 \\ M = 8
 % & OPTICS & 0.110 & 0.359 & 21 \\
 & OPTICS & {\bf 0.280} & {\bf 0.459} & 3 \\
 & SpectACl & 0.252 & 0.381 & 7 \\
\hline \multirow{5}{4em}
{Cancer}
 & MSE & {\bf 0.743} & {\bf 0.628}  & 2 \\ 
 & Spectral & 0.583 & 0.487  & 2 \\
 & $K$-means & 0.491 & 0.465 & 2 \\
 & HDBSCAN & 0.625 & 0.487  & 4 \\
 & HDBSCAN (2) & 0.000 & 0.000  & 1 \\
 %& OPTICS & 0.737 & 0.620 & 2 \\
 & OPTICS & 0.737 & 0.620 & 2 \\
 & SpectACl & 0.707 & 0.586 & 2 \\
 \hline \multirow{5}{4em}
{Digits}
 & MSE & {\bf 0.864} & {\bf 0.898}   & 10 \\ 
 & Spectral & 0.781 & 0.892  & 10 \\
 & $K$-means & 0.615 & 0.731 & 10 \\
 & HDBSCAN & 0.575 & 0.770  & 22 \\
 & HDBSCAN (2) & 0.559 & 0.762  & 9 \\
 %& OPTICS & 0.259 & 0.628 & 10 \\
 & OPTICS & 0.585 & 0.777 & 8 \\
 & SpectACl & 0.564 & 0.750 & 10 \\
\hline \multirow{5}
{4em}
{Letters}
 & MSE & {\bf 0.193} &  0.447   & 26 \\ 
 & Spectral & 0.098 & 0.408  & 26 \\
 & $K$-means & 0.130 & 0.356 & 26 \\
 & HDBSCAN & 0.023 & {\bf 0.536}  & 463 \\
 & HDBSCAN (2) & 0.003 & 0.064  & 3 \\
 %& OPTICS & 0.047 & 0.343 & 24 \\
 & OPTICS & 0.058 & 0.392 & 46 \\
 & SpectACl & 0.092 & 0.264 & 26 \\ 
 \hline \multirow{5}{4em}
{MNIST}
 & MSE & {\bf 0.854} & {\bf 0.854}   & 10 \\ 
 & Spectral & 0.632 & 0.746  & 10 \\
 & $K$-means & 0.656 & 0.751 & 10 \\
 & HDBSCAN & 0.712 & 0.795  & 11 \\
 & HDBSCAN (2) & 0.603 & 0.766  & 8 \\
 %& OPTICS & 0.444 & 0.689 & 11 \\
 & OPTICS & 0.623 & 0.770 & 8 \\
 & SpectACl & 0.817 & 0.831 & 10 \\
\hline
\end{tabular}
%}
\end{center}
\end{table}

% \begin{table} 
% \begin{center}
% \begin{tabular}{ |c|c|c|c|c| } 
%  \hline
% dataset & algorithm  & ARI & NMI & $|\hat{C}|$ \\
% \hline \multirow{5}{4em}
% {Digits}
%  & MSE & {\bf 0.864} & {\bf 0.898}   & 10 \\ 
%  & Spectral & 0.781 & 0.892  & 10 \\
%  & $K$-means & 0.615 & 0.731 & 10 \\
%  & HDBSCAN & 0.575 & 0.770  & 22 \\
%  & HDBSCAN (2) & 0.559 & 0.762  & 9 \\
%  %& OPTICS & 0.259 & 0.628 & 10 \\
%  & OPTICS & 0.585 & 0.777 & 8 \\
%  & spectACl & 0.564 & 0.750 & 10 \\
% \hline \multirow{5}
% {4em}
% {Letters}
%  & MSE & {\bf 0.193} &  0.447   & 26 \\ 
%  & Spectral & 0.098 & 0.408  & 26 \\
%  & $K$-means & 0.130 & 0.356 & 26 \\
%  & HDBSCAN & 0.023 & {\bf 0.536}  & 463 \\
%  & HDBSCAN (2) & 0.003 & 0.064  & 3 \\
%  %& OPTICS & 0.047 & 0.343 & 24 \\
%  & OPTICS & 0.058 & 0.392 & 46 \\
%  & spectACl & 0.092 & 0.264 & 26 \\ 
%  \hline \multirow{5}{4em}
% {MNIST}
%  & MSE & {\bf 0.854} & {\bf 0.854}   & 10 \\ 
%  & Spectral & 0.632 & 0.746  & 10 \\
%  & $K$-means & 0.656 & 0.751 & 10 \\
%  & HDBSCAN & 0.712 & 0.795  & 11 \\
%  & HDBSCAN (2) & 0.603 & 0.766  & 8 \\
%  %& OPTICS & 0.444 & 0.689 & 11 \\
%  & OPTICS & 0.623 & 0.770 & 8 \\
%  & spectACl & 0.817 & 0.831 & 10 \\
% \hline
% \end{tabular}
% \caption{ \label{table:benchmark-results-sweep2} 
% $|\hat{C}|$ is the number of clusters output by the algorithm.  For HDBSCAN and OPTICS, the set of noise points counts as a cluster.
% %For the MNIST dataset, we used TSNE to reduce the dimensionality to 2. 
%  }
% \end{center}
% \end{table}

\begin{table} 
\begin{center}
\caption{ \label{table:benchmark-results-auto} 
$|\hat{C}|$ is the number of clusters output by the algorithm.
 }
\begin{tabular}{ |c|c|c|c|c| } 
 \hline
dataset & algorithm  & ARI & NMI & $|\hat{C}|$ \\
 \hline \multirow{2}{4em}{Iris}
 & MSE & {\bf 0.886} & {\bf 0.871} & 3\\ 
 & MSE (auto) &  0.835 &  0.833 & 3\\ 
 \hline \multirow{2}{4em}{Wine} 
 & MSE & {\bf 0.439} & {\bf 0.430}  & 3\\ 
 & MSE (auto) &  0.359 &  0.420 & 3\\ 
 \hline \multirow{2}{4em}{Seeds}
 & MSE &  0.725 & 0.682  & 3 \\ 
 & MSE (auto) &  0.725 & 0.682  & 3 \\ 
 \hline \multirow{2}{4em}{Glass}
 & MSE & 0.232 & 0.379  & 7 \\
 & MSE (auto) & 0.232 & 0.379  & 7 \\
\hline \multirow{2}{4em}
{Cancer}
 & MSE & {\bf 0.743} & {\bf 0.628}  & 2 \\ 
 & MSE (auto) &  0.694 &  0.595 & 2\\ 
\hline \multirow{2}{4em}
{Digits}
 & MSE & {\bf 0.864} & {\bf 0.898}   & 10 \\
 & MSE (auto) & 0.819 &  0.879 & 10 \\ 
 \hline \multirow{2}{4em}
{Letters}
 & MSE &  0.193 & 0.447  & 26 \\
& MSE (auto) &  0.193 & 0.447  & 26 \\ 
 \hline 
\end{tabular}
\end{center}
\end{table}

For the other algorithms tested, we used grid search over the parameters to maximize ARI.  
For spectral clustering, we tried all $K_n$ in the range $\{1,2,...,20\}$ and reported the clustering with the best ARI. For HDBSCAN, we tried all $N_p$ in the range $\{2,3, ..., 20\}$ and reported the clustering with best ARI.
For HDBSCAN (2), we used the same $M$ and $N_p$ we used for MSE.
For OPTICS, we used the same $N_p$ value that we used for MSE, and tried all $X_i$ values in the set $\{0, 0.05, 0.1, ..., 0.95\}$ and all $M$ values in the set $\{\delta \cdot |X| / K : \delta \in \Delta\}$ where $\Delta = \{0.05, 0.10, 0.15, ..., 1\}$,  and reported the clustering with best ARI.  For MNIST, we tested $\Delta = \{0.1, 0.2, 0.3, ..., 1\}$ for OPTICS to save time.
For SpectACl, we tried all $\epsilon$ values in the range $\{0, 0.1, 0.2, ..., 100\}$.  For some large datasets where the optimal $\epsilon$ was clearly less than 20, we tried all $\epsilon$ values in the range $\{0, 0.1, 0.2, ..., 20\}$.

For MSE, and HDBSCAN (2), $N_p$ was set to 3 for all datasets and $M$ was set to 35 for Iris, 30 for Wine, 50 for Seeds, 3 %8
for Glass, 100 for Cancer and Digits, 70 for Letters, and 600 for MNIST.
For MSE, $D$ was set to 20 for Iris, Wine, Glass, and Cancer, 2 for Seeds, Letters and MNIST, and 1.5 for Digits.
For spectral clustering, in order of dataset appearance in Table \ref{table:benchmark-results-sweep}, the chosen $K_n$ values are $4, 6, 18, 3, 5, 4, 19, 11$.
For HDBSCAN, in order of appearance in Table \ref{table:benchmark-results-sweep}, the chosen $N_p$ values are $3, 20, 6, 3, 6, 4, 3, 14$.  
For OPTICS, in order of appearance in Table \ref{table:benchmark-results-sweep}, the chosen $M$ values are $38, 47, 56, 20, 57, 162, 54, 800$, and 
the chosen $X_i$ values are $0,$ $0,$ $0,$ $0.15,$ $0.05,$ $0,$ $0,$ $0.05$. 
%$0, 0.25, 0, 0.2, 0.05, 0, 0, 0$.  for when M is same as in MSE
For SpectACl, in order of appearance in Table \ref{table:benchmark-results-sweep}, the chosen $\epsilon$ values are $3.3,$ $44.3,$ $2.0,$ $2.3,$ $77.2,$ $33.2,$ $8.9,$ $9.9$.  

The second set of results for benchmark datasets is reported in Table \ref{table:benchmark-results-auto}.  MSE (auto) refers to MSE where $M$ values and $D$ values are optimized to give the best Calinski-Harabasz score, which is an internal clustering metric, meaning that it does not use the ground truth labels to asses clustering quality.  
For all of these experiments, we set $N_p = 3$. 
We chose the $M$ from $\{\delta \cdot |X|/K \; : \; \delta \in \Delta \}$ where $\Delta = [0.025, 0.05, 0.075, ..., 0.975]$ and chose $D$ from $[1.5, 2, 20]$.  We chose from these values of $M$ because we know the number of clusters $K$, and we choose from these values of $D$ because we have observed at least one of these values to work well on a wide range of datasets.  In Table \ref{table:benchmark-results-auto}, we also reported the statistics for the MSE clusterings from Table \ref{table:benchmark-results-sweep} for comparison.  We observe that in general, the decrease in clustering quality is not large on these datasets.
The average decrease in ARI is 0.032, and the average decrease in NMI is 0.014.  The average percent decrease in ARI is 5.1\% and the average percent decrease in NMI is 2.0\%. 
We did not test on MNIST due to the slower speed of the algorithm on this dataset.
Note that we only tested this approach for this one internal clustering quality measure, and the results may improve if another internal clustering quality measure (e.g.  DBCV score \cite{Moulavi2014}) is used to optimize the clustering.  Additionally, $N_p$ can also be optimized using this approach.
%such as the DBCV score \cite{Moulavi2014} (or other cluster quality score)

% In Table \ref{table:simulation-results-sweep}, we also compare the same algorithms for 5 artificial datasets created using the Python scikit-learn library which were used in the example titled ``Comparing different clustering algorithms on toy datasets'' on the scikit-learn website \cite{scikitlearn_toy_datasets}.  
% In Table \ref{table:simulation-results-sweep}, we compare the algorithms on the 5 artificial datasets \cite{scikitlearn_toy_datasets}.

The results for artificial datasets in Figures \ref{fig:artificial_datasets} are reported in Table \ref{table:simulation-results-sweep}.
For MSE and HDBSCAN (2), $N_p$ was set to 3 for all datasets, and $M$ was set to 60 for all datasets in Table \ref{table:simulation-results-sweep}.
For MSE, $D$ was set to 10 for Varied Variance Blobs, and 2 for all other datasets in Table \ref{table:simulation-results-sweep}.
For spectral clustering, in order of appearance in Table \ref{table:simulation-results-sweep}, the chosen $K_n$ values are $5, 5, 18, 6, 15$.
For HDBSCAN, in order of appearance in Table \ref{table:simulation-results-sweep}, the chosen $N_p$ values are $ 3, 6, 9, 9, 14$. 
For OPTICS, $N_p$ was set to 3, and in order of appearance in Table \ref{table:simulation-results-sweep}, the chosen $M$ values are $225, 225, 133, 167, 167$, and the chosen $X_i$ values are $0.0, 0.0, 0.05, 0.0, 0.1$
%$0.15, 0.1, 0.3, 0.2, 0.15$.  
For SpectACl, in order of appearance in Table \ref{table:simulation-results-sweep}, the chosen $\epsilon$ values are $0.2, 0.2, 1.2, 0.8, 1.3$.

\begin{table} 
\begin{center}
\caption{ \label{table:artificial-datasets} artificial datasets and their properties}
\begin{tabular}{ |c|c|c|c|c| } 
 \hline
dataset & \# points  & \# features & \# clusters & min. cluster size\\
\hline
 Two Circles & 500 & 2 & 2 & 250 \\
 \hline
 Two Moons & 500 & 2 & 2 & 250 \\
 \hline
 Fixed Var. Blobs & 500 & 2 & 3 & 166 \\
 \hline
 Anisotropic & 500 & 2 & 3 & 166 \\
 \hline
 Varied Var. Blobs & 500 & 2 & 3 & 166 \\
 \hline
\end{tabular}
\end{center}
\end{table}

\begin{table} 
\begin{center}
\caption{
$|\hat{C}|$ is the number of clusters output by the algorithm.  For HDBSCAN and OPTICS, the set of noise points counts as a cluster.
%For the MNIST dataset, we used TSNE to reduce the dimensionality to 2. 
 }
\label{table:simulation-results-sweep} 
\begin{tabular}{ |c|c|c|c|c| } 
 \hline
dataset & algorithm  & ARI & NMI & $|\hat{C}|$ \\
 \hline \multirow{6}{5em}{Two Circles}
 & MSE & 1.000 & 1.000 & 2\\ 
 & Spectral & 1.000 & 1.000 & 2\\
 & $K$-means & -0.002 & 0.000 & 2 \\
 & HDBSCAN & 1.000 & 1.000  & 2 \\
 & HDBSCAN (2) & 1.000 & 1.000  & 2 \\
 % & OPTICS & 1.000 & 1.000 & 2 \\
 & OPTICS & 1.000 & 1.000 & 2 \\
 & SpectACl & 1.000 & 1.000 & 2 \\
 \hline \multirow{6}{5em}{Two Moons} 
 & MSE & 1.000 & 1.000  & 2\\ 
 & Spectral & 1.000 & 1.000 & 2 \\ 
 & $K$-means & 0.233 & 0.176 & 2 \\
 & HDBSCAN & 1.000 & 1.000  & 2\\ 
 & HDBSCAN (2) & 0.699 & 0.737  & 4 \\ 
 % & OPTICS & 1.000 & 1.000 & 2 \\ 
 & OPTICS & 1.000 & 1.000 & 2 \\ 
 & SpectACl & 1.000 & 1.000 & 2 \\
 \hline \multirow{6}{5em}{Fixed Variance Blobs}
 & MSE & 0.964 & 0.948  & 3 \\ 
 & Spectral & {\bf 0.976} & {\bf 0.961}  & 3 \\ 
 & $K$-means & 0.970 & 0.954 & 3 \\
 & HDBSCAN & 0.867 & 0.847  & 4 \\
 & HDBSCAN (2) & 0.568 & 0.729  & 3 \\
 % & OPTICS & 0.569 & 0.733 & 2 \\
 & OPTICS & 0.726 & 0.734 & 3 \\
 & SpectACl & 0.964 & 0.942 & 3 \\
 \hline \multirow{6}{5em}{Anisotropic}
 & MSE & {\bf 1.000} & {\bf 1.000}  & 3 \\ 
 & Spectral & 0.994 & 0.989  & 3  \\
 & $K$-means & 0.555 & 0.593 & 3 \\
 & HDBSCAN & 0.917 & 0.888  & 4  \\
 & HDBSCAN (2) & 0.991 & 0.983 & 4 \\
 % & OPTICS & 0.970 & 0.951 & 4 \\
 & OPTICS & {\bf 1.000} & {\bf 1.000} & 3 \\
 & SpectACl & 0.994 & 0.989 & 3 \\
\hline \multirow{6}{5em}
{Varied Variance Blobs}
 & MSE & 0.896 & 0.867  & 3 \\ 
 & Spectral & 0.896 & 0.873  & 3 \\
 & $K$-means & 0.787 & 0.778 & 3 \\
 & HDBSCAN & 0.808 & 0.809  & 4 \\
 & HDBSCAN (2) & 0.844 & 0.817  & 4 \\
 % & OPTICS & 0.906 & 0.865 & 3 \\
 & OPTICS & 0.924 & 0.894 & 3 \\
 & SpectACl & {\bf 0.930} & {\bf 0.897} & 3 \\
\hline 
\end{tabular}
\end{center}
\end{table}

% Spectral: 5, 5, 18, 6, 15

% HDBSCAN (d): 3, 6, 9, 9, 14 

% Optics $X_i$:  0.15   0.1  0.3   0.2   0.15

\begin{figure}%[htp]
\centering

    \begin{subfigure}{\includegraphics[width=0.45\linewidth]{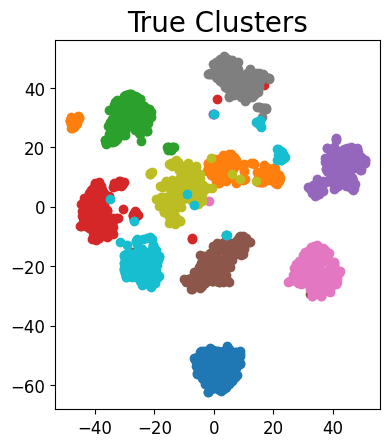}}
    \end{subfigure}
\hfill
    \begin{subfigure}{\includegraphics[width=0.45\linewidth]{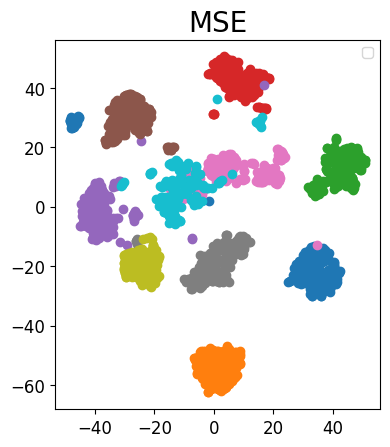}}
    \end{subfigure}
\hfill
    \begin{subfigure}{\includegraphics[width=0.45\linewidth]{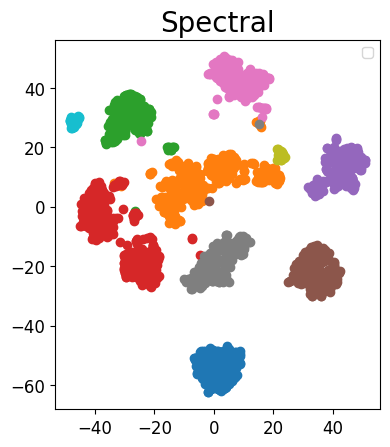}}
    \end{subfigure}
\hfill
    \begin{subfigure}{\includegraphics[width=0.45\linewidth]{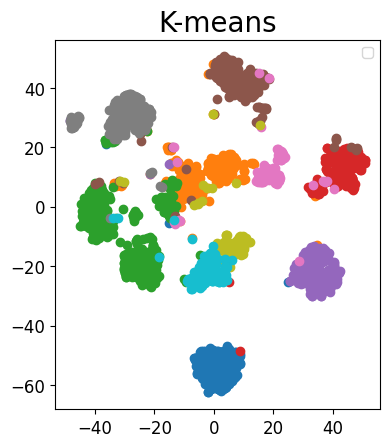}}
    \end{subfigure}
% \hfill
%     \begin{subfigure}{\includegraphics[width=0.45\linewidth]{images/Digits_TSNE_HDBSCAN_d.png}}
%     \end{subfigure}
% \hfill
%     \begin{subfigure}{\includegraphics[width=0.45\linewidth]{images/Digits_TSNE_HDBSCAN.png}}
%     \end{subfigure}
% \hfill
%     \begin{subfigure}{\includegraphics[width=0.45\linewidth]{images/Digits_TSNE_OPTICS.png}}
%     \end{subfigure}

\caption{t-SNE plots for the Digits dataset.}

\label{fig:Digits_TSNE}

\end{figure}

Observe that with a lower bound on the minimum cluster size ($M$), and some knowledge of the maximum difference in maximum density among different clusters ($D)$, MSE gives better ARI than the other algorithms on all %artificial and 
benchmark datasets except Glass, and gives competitive ARI for all artificial datasets.  On the Glass dataset, OPTICS gives the best ARI, but outputs an incorrect number of clusters. The NMI values obtained by our algorithm are also competitive.   
% MSE performs strongly on the artificial datasets as well.
Compared to the other algorithms, MSE performs well more consistently across the datasets tested, despite the fact that the parameters were not optimized in an exhaustive manner.
As an example, in the case of the Digits dataset, it is clear in Figure \ref{fig:Digits_TSNE} that our algorithm does a significantly better job of recovering the Digits clustering than the other algorithms.
%While %the focus of this paper is not computational complexity, and 
While our implementation of MSE was not optimized for speed, MSE clustered the MNIST dataset in three minutes on a laptop computer with 32 threads and 64 GB of RAM.  MSE ran much faster on the other datasets.

% in terms of ARI in all cases except the Glass dataset (on which it performs similarly), substantially better than HDBSCAN in all cases except the Glass dataset (on which HDBSCAN outputs the incorrect number of clusters), and substantailly better than OPTICS on all datasets.  The NMI values obtained by our algorithm are also competitive.  

% For MSE, $D$ (and similarly $M$ and $N_p$) can be optimized by running MSE for various values of $D$ and using the value that gives the clustering with the best DBCV score \cite{Moulavi2014} (or other cluster quality score). %though we leave this optimization for future work. 

\section{Conclusion}

In this work, we propose an
information-theoretic characterization of when a $K$-clustering is ambiguous, and design an algorithm called Minimal Seed Expansion (MSE) that provably recovers the clustering
whenever it is unambiguous. This characterization formalizes the
situation when two high density regions within a cluster are
separable enough that they look more like two distinct clusters
than two truly distinct clusters in the clustering.
We then implement and test a version of MSE that is modified to effectively handle overlapping clusters, and observe that it displays improved performance on many datasets without its parameters exhaustively optimized for each dataset tested. This improvement is in comparison to widely used algorithms for non-convex cluster recovery whose parameters are optimized using grid search independently for every dataset tested.  MSE also performs well more consistently than all other algorithms tested in these experiments.  We also optimized the parameters for MSE using grid search to maximize the Calinski-Harabasz score (an internal clustering quality measure), and observed little decrease in ARI and NMI compared to the case where parameters were chosen to give competitive ARI.  This suggests that MSE can also be used effectively without manual parameter tuning.

\section{Acknowledgments}
The work of K.M. and I.S. was supported in part by the National Science Foundation CAREER Award under Grant CCF-2046991.  The authors thank the anonymous reviewers of ISIT 2025 for their helpful comments.

\bibliographystyle{ieeetr}
\bibliography{refs}

\section{Appendix}

\subsection{Strong Separability}

A similar, yet stronger condition than weak separability %that we later prove is a sufficient condition for clustering recoverability 
is what we call strong separability.  
%For a cluster $c \in C$, let $X_c^*$ denote the set of all points $x$ in $c$ such that $\epsilon_{N_p}(x) = \min_{y \in c} \epsilon_{N_p}(y)$.  In other words, $X_c^*$ denotes the set of all points $x$ in $c$ that have highest density among points in $c$.

\begin{definition}
    For a given $N_p$, $C$ is called strongly separable if there exists some $A \in \mathbb{R}$ such that one of the following equivalent conditions holds.
    \begin{enumerate} %[noitemsep, topsep=0.2pt]
    \item For each $c \in C$, 
    %$\alpha(c) \leq A$,
    $c$ is $(A \cdot \min_{x \in c} \epsilon_{N_p}(x))$-connected, and $A \cdot \min_{x \in c} \epsilon_{N_p}(x) < \min_{c'\in C, \; c' \neq c }\epsilon(c, c')$.  
    \item For each $c \in C$, $c^*(x, A \cdot \epsilon_{N_p}(x)) = c$ for any $x \in X_c^*$
    \end{enumerate}
    The minimum such $A$ is denoted by $A^*(C)$.
\end{definition}

%Intuitively, the definition says that for a clustering $C$, there exists a constant $A^*(C)$ such that for each cluster $c$, all points in the cluster are $\epsilon$-connected where $\epsilon$ is a constant $A^*(C)$ times the sparsity of the highest density point in $c$, but no other cluster is $\epsilon$-connected to $c$.  
The definition specifies for each cluster $c$, an $\epsilon$ relative to the sparsity of the maximum density point in $c$, such that no cluster can be $\epsilon$-connected to $c$, yet all points in $c$ must be $\epsilon$-connected.
% It therefore naturally includes clusterings with arbitrary differences in density among the points within each cluster, and arbitrary difference in average density among different clusters, so long as the clusters are separated enough relative to the sparsity values of their respective maximum density points.  
The condition is therefore naturally satisfied by clusterings that have arbitrarily shaped
clusters with arbitrarily many relatively separated
regions of high density and arbitrary variation in density among different clusters, 
%arbitrary variation in density within clusters,  
so long as the clusters are separated enough relative to the sparsity values of their respective maximum density points.  Unlike weak separability, if $C$ is strongly separable for $N_p$, then it is the unique strongly separable clustering for $N_p$ and can be found efficiently by Theorem \ref{thm:strong_sep_main}, which follows immediately from Theorem \ref{thm:LM_sep_main} and Lemmas \ref{lem:strong_to_weak} and \ref{lem:strong-sep-implies_ell-sep}.
\begin{theorem} \label{thm:strong_sep_main}
    If $C$ is strongly separable for a given $N_p$, then $C$ is the unique strongly separable clustering for $N_p$, and can be found in $O(|X|^3 \log(|X|))$ time.
\end{theorem}
% \begin{proof}
%     This follows from Theorem \ref{thm:LM_sep_main} and Lemma \ref{lem:strong-sep-implies_ell-sep}.
% \end{proof}

Perhaps a more intuitive condition that implies strong separability is given in the following lemma, which for each cluster $c \in C$, bounds the maximum variation in density among points in $C$.
Let \[\alpha(c) = \max_{x \in c} \epsilon_{N_p}(x) / \min_{y \in c} \epsilon_{N_p}(y)\]
%$\alpha(c) = \frac{\max_{x \in c} \epsilon_{N_p}(x)}{\min_{y \in c} \epsilon_{N_p}(y)}$
be the ratio of the sparsity of the minimum density point in $X$ to that of the maximum density point in $X$.
\begin{lemma}
    For a given $N_p$, if there exists some $A \in \mathbb{R}$ such that for every $c \in C$, $\alpha(c) \leq A$, $c$ is $(\max_{x \in c} \epsilon_{N_p}(x))$-connected, and $A \cdot \min_{x \in c} \epsilon_{N_p}(x) < \min_{c'\in C, \; c' \neq c }\epsilon(c, c')$, 
    then $C$ is strongly separable.
\end{lemma}
\begin{IEEEproof}
    The fact that for each $c \in C$, $\alpha(c) \leq A$ and $c$ is $(\max_{x \in c} \epsilon_{N_p}(x))$-connected implies that for each $c \in C$, $c$ is $(A \cdot \min_{x \in c} \epsilon_{N_p}(x))$-connected.
\end{IEEEproof}

If $C$ is strongly separable, then it is weakly separable as proved in Lemma \ref{lem:strong_to_weak}. However, strong separability of $C$ does not imply that $C$ is the unique weakly separable $|C|$-clustering for $N_p$. Furthermore, there exist weakly separable clusterings that are not strongly separable.  For example, for $N_p=2$, $C = [\{1,3,5\}, \{8, 10, 11, 13\}]$
% \begin{align*}
% C = [\{1,3,5\}, \{8, 10, 11, 13\}]
% \end{align*}
is weakly separable but not strongly separable since the  points in the second cluster imply that $A^*(C) \geq 2$, but $c^*(3, 2\cdot \epsilon_{N_p}(3)) = c^*(3, 4) = X$. %includes the all points in $X$.
%In this example, 
Intuitively, $C$ is the correct $2$-clustering, thus showing that strong separability as a sufficient condition for clustering recovery is still not general enough. %For this reason, 
Therefore, we introduce LM-separability as a more inclusive sufficient condition for recoverability.  Figure \ref{fig:weak_not_strong} gives a larger example of a weakly separable clustering that is not strongly separable.
%for $N_p = 5$.

\begin{lemma}
\label{lem:strong_to_weak}
    For a given $N_p$, if $C$ is strongly separable, then it is weakly separable.
\end{lemma}
\begin{IEEEproof}
    This follows immediately from the first equivalent definition of strong separability.
\end{IEEEproof}

LM-separability is more natural than strong separability because LM-separability and weak separability hold precisely when a clustering is unambiguous.  In fact, strong separability implies LM-separability.
%by Lemma \ref{lem:strong-sep-implies_ell-sep}.

\begin{lemma} \label{lem:strong-sep-implies_ell-sep}
    For a given $N_p$, if $C$ is strongly separable, then it is LM-separable.
\end{lemma}
\begin{IEEEproof}
    If there is only one local maximum per cluster, then this holds trivially.  %Suppose $C$ contains at least one cluster that has at least two local maximums,
    Suppose $C$ is strongly separable but there exists a local maximum $x$ such that 
    $\min_{y \in c(x): \; \epsilon_{N_p}(y) < \epsilon_{N_p}(x)} A(x, y) \geq \min_{c, c' \in C} \min_{z \in X_c^*} A(z, c').$
    % \begin{equation*}
    %     \min_{y \in c(x): \; \epsilon_{N_p}(y) < \epsilon_{N_p}(x)} A(x, y) \geq \min_{c, c' \in C} \min_{z \in X_c^*} A(z, c').
    % \end{equation*}
    Since $\epsilon_{N_p}(x)  > \epsilon_{N_p}(w)$ for all $w \in X_{c(x)}^*,$ this implies that
    % \begin{align*}
    % A(w, x) \geq \min_{c, c' \in C} \min_{z \in X_c^*} A(z, c')
    % \end{align*}
    \begin{align} A(w, x) \geq \min_{c, c' \in C} \min_{z \in X_c^*} A(z, c') \nonumber
    \end{align}
    for all $w \in X_{c(x)}^*$,
    % \begin{align}
    % \min_{ A : \; 
    % x \in c^*(x_{c(x)}^*, A \cdot \epsilon_{N_p}(x_{c(x)}^*))} A \geq \min_{c \in C} \min_{c' \in C, c' \neq c} A(x_c^*, c'),
    % \end{align}
    which in turn implies that $A^*(C) \geq\min_{c, c' \in C} \min_{z \in X_c^*} A(z, c')$.
    Thus, for the $c, c', z$ that minimize $\min_{c, c' \in C}  \min_{z \in X_c^*} A(z, c')$, we have that $c^*(z, \; A^*(C)\cdot \epsilon_{N_p}(z))$ contains at least one point from $c'$. 
    This is a contradiction to strong separability.   
\end{IEEEproof}

Furthermore, weak separability together with LM-separability does not imply strong separability.
Let $N_p = 2$, and consider the clustering $C = [\{7, 8, 10, 13\}, \{17, 19, 21\}]$.  $C$ is weakly separable, and is LM-separable as $7,8,17,19,21$ are the only local maxima.  $C$ is not strongly separable because the first cluster implies that $A^*(C) \geq 3$, but $c^*(19, 3 \cdot \epsilon_{N_p}(19)) = c^*(19, 6) = X$.   Figure \ref{fig:weak_not_strong} shows a larger example of a clustering that is not strongly separable, but is weakly separable and LM-separable, therefore guaranteeing recoverability by Theorem \ref{thm:LM_sep_main}.

For a given $N_p$, it is in general not possible to recover the strongly separable clustering by simply finding an extension of the (partial) clustering given by the $\epsilon$-cut of the dendrogram that contains $K$ clusters (if there exists such an $\epsilon$)
as proved in Lemma \ref{lem:dbscan_insufficient_strongly}.  Since DBSCAN follows this approach, it is not sufficient for finding the strongly separable clustering.
%Intuitively, this is because a strongly separable clustering may have a high variation in density among different clusters that an $\epsilon$-cut cannot capture.   
\begin{lemma} \label{lem:dbscan_insufficient_strongly}
    There exists a strongly separable clustering for some $N_p$ that does not extend any (partial) clustering given by an $\epsilon$-cut of the dendrogram.
\end{lemma} 
\begin{IEEEproof}
    Recall that there can only be one $\epsilon$-cut of the dendrogram that gives a (partial) $K$-clustering.  Let $N_p = 3$, and suppose that 
    $C = [\{1, 3, 5, 7.02, 9.02, 11.02\},$ $\{17, 18, 19, 20\},$ 
         $\{22.01, 23.01, 24.01, 25.01\}].$
    % \begin{align*}
    %     C = [\{1, 3, 5, 7.02, 9.02, 11.02\}, \{17, 18, 19, 20\}, 
    %      \{22.01, 23.01, 24.01, 25.01\}].
    % \end{align*}
    % \begin{align*} C = [&\{1, 3, 5, 7.02, 9.02, 11.02\}, \{17, 18, 19, 20\}, 
    %     \\ & \{22.01, 23.01, 24.01, 25.01\}]. \end{align*}
    This is a strongly separable clustering and $A^*(C) = 2$.  The only $\epsilon$-cut that gives a (partial) clustering with three clusters is chosen by setting $\epsilon = 2.01$, and is given by 
    \begin{align*}
    C' = [\{3\}, \{9.02\},  
         \{17, 18, 19, 20, 22.01, 23.01, 24.01, 25.01\}].
    \end{align*}
    % \begin{align*}
    %     C' = [&\{3\}, \{9.02\},
    %     \\ & \{17, 18, 19, 20, 22.01, 23.01, 24.01, 25.01\}].
    % \end{align*} 
    The only weakly separable clustering that is an extension of $C'$ is 
    $[\{1, 3, 5\},$ $\{7.02, 9.02, 11.02\},$ $\{17, 18, 19, 20, 22.01, 23.01, 24.01, 25.01\}].$
    % \begin{align*}
    %     [&\{1, 3, 5\}, \{7.02, 9.02, 11.02\},
    %     \\ & \{17, 18, 19, 20, 22.01, 23.01, 24.01, 25.01\}].
    % \end{align*} 
    %Thus, the partial $K$-clustering given by an $\epsilon$-cut of the dendrogram cannot be extended to a strongly separable clustering.
\end{IEEEproof}
 
%In the clustering $C$ used to prove Lemma \ref{lem:dbscan_insufficient}, the distance separating the points $1, 3, 5$ from the points $7.02, 9.02, 11.02$ within the first cluster is $2.02$, which is larger than the distance of $2.01$ separating the clusters $\{17, 18, 19, 20\}$ and $\{22.01, 23.01, 24.01, 25.01\}$. However, $2.02$ is very similar to the distance of $2$ separating the other pairs of adjacent points in the first cluster, while $2.01$ is very large compared to the distance of $1$ separating the adjacent points in the second cluster.  $C$ is therefore a more natural clustering than $C'$ in a sense, but an $\epsilon$-cut is unable to capture $C$ because it only considers absolute distances when separating clusters, without taking cluster density into account.

\end{document}